\documentclass[5p,times]{elsarticle}

\usepackage{amssymb}
\usepackage{amsmath}

\usepackage{graphicx}
\usepackage{xcolor}
\usepackage{dcolumn}
\usepackage{bm}
\usepackage{graphicx}
\usepackage{overpic}
\usepackage{dcolumn}
\usepackage{bm}
\usepackage{epstopdf}

\usepackage[utf8]{inputenc}
\usepackage{amsmath}
\usepackage[T1]{fontenc}
\usepackage{booktabs, array, mathptmx, float, tabularx, booktabs, lipsum, multirow}
\usepackage{siunitx, xcolor}

\graphicspath{{figs/}} 

\usepackage[colorlinks,linkcolor=blue,anchorcolor=blue,citecolor=blue]{hyperref}
\usepackage[justification=justified]{caption}
\usepackage{graphicx}
\usepackage{xcolor}
\usepackage{dcolumn}
\usepackage{bm}
\usepackage{graphicx}
\usepackage{overpic}
\usepackage{dcolumn}
\usepackage{bm}
\usepackage{epstopdf}

\usepackage[utf8]{inputenc}
\usepackage{amsmath}
\usepackage[T1]{fontenc}
\usepackage{booktabs, array, mathptmx, float, tabularx, booktabs, lipsum, multirow}
\usepackage{siunitx, xcolor}

\usepackage[colorlinks,linkcolor=blue,anchorcolor=blue,citecolor=blue]{hyperref}
\usepackage[justification=justified]{caption}

\usepackage[english]{babel}
\usepackage{microtype}

\journal{Physics Letter B}

\begin{document}

\begin{frontmatter}



\title{Probing vacuum birefringence in an Ultrastrong Laser Field\\
via High-energy Gamma-ray Polarimetry}



\author[xjtu]{Da-Lin Wang}
\ead{wangdalin133@stu.xjtu.edu.cn}

\author[xjtu]{Xian-Zhang Wu}
\ead{3123303391@stu.xjtu.edu.cn}

\author[xjtu]{Rui-Qi Qin}
\ead{qrq2833757055@stu.xjtu.edu.cn}

\author[tsinghua]{Jiang-Tao Han}
\ead{hanjt25@mails.tsinghua.edu.cn}

\author[xjtu]{Peng-Pei Xie}
\ead{xiepengpei@stu.xjtu.edu.cn}

\author[xjtu]{Bing-Jun Li\corref{cor1}}
\ead{lbj_xjtu@stu.xjtu.edu.cn}

\author[zhengzhou]{Huai-Hang Song}
\ead{huaihangsong@gmail.com}

\author[xjtu]{Yan-Fei Li\corref{cor2}}
\ead{liyanfei@xjtu.edu.cn}

\affiliation[xjtu]{organization={School of Energy and Power Engineering, Xi'an Jiaotong University},   
            city={Xi'an},
            postcode={710049},
            country={China}}

\affiliation[tsinghua]{organization={Institute of Nuclear and New Energy Technology, Tsinghua University},
            city={Beijing},
            postcode={100084},
            country={China}}

\affiliation[zhengzhou]{organization={Laboratory of Zhongyuan Light, School of Physics, Zhengzhou University},
            city={Zhengzhou},
            postcode={450001},
            country={China}}

\cortext[cor1]{Corresponding author}
\cortext[cor2]{Principle Corresponding author}

\begin{abstract}

Vacuum birefringence (VB), a fundamental prediction of nonlinear QED, has eluded direct laboratory detection due to its extreme weakness. We propose a compact, “self-probing” scheme where a GeV electron beam collides head-on with a petawatt laser pulse. Circularly polarized $\gamma$-ray photons, generated via nonlinear Compton scattering in the same pulse, then probe the birefringent vacuum it induces. 
This integrated design bypasses the stringent synchronization and beam transport requirements of traditional pump–probe setups. Our nonperturbative strong-field QED simulations reveal a clear VB signature: conversion of circular to linear polarization, with the induced Stokes parameter $S_1$ reaching $\sim 0.019$ within the selected angular range. 
This corresponds to a refractive index difference $\Delta n = 1.829 \times 10^{-4}$ over micron-scale paths, directly measurable as a high-contrast “X-shape” asymmetry in $e^+e^-$ pair distributions. The scheme provides a feasible path to first laboratory VB detection with current laser and accelerator technologies.

\end{abstract}



\begin{keyword}
vacuum birefringence\sep nonlinear Compton scattering \sep photon polarization


\end{keyword}

\end{frontmatter}



\section{Introduction}
\label{sec1}


Vacuum birefringence (VB), a phenomenon where the quantum vacuum acquires different refractive indices for distinct light polarizations in a strong field, is a fundamental prediction of nonlinear quantum electrodynamics (QED). 
It originates from vacuum polarization (VP), the transient fluctuation of virtual electron-positron pairs induced by an intense electromagnetic background\cite{dunne2005heisenberg, FolgerungenAusDiracschen1936, euler1935streuung}. 
While theoretically well-established, the experimental detection of VB remains a major challenge. The effect is governed by the quantum nonlinearity parameter $\chi_\gamma$ and is exceedingly weak at field strengths achievable in laboratories, which are far below the critical Schwinger field ($E_{\mathrm{cr}} \approx 1.3 \times 10^{18} \,\mathrm{V/m}$). 
Consequently, amplifying the feeble VB signal to a measurable level demands innovative experimental strategies.

A common strategy in these attempts has been the pump--probe architecture, which employs separate systems to generate the strong field and to produce or probe the photons. These approaches can be categorized by the probe photon energy. First, optical-wavelength experiments such as PVLAS and BMV employ high-finesse cavities and static magnetic fields to achieve extraordinary sensitivity~\cite{Zavattini2016, Cadene2013,Ejlli2020}. 
Despite their high sensitivity, they are fundamentally limited by the low photon energy and weak field strength, struggling to overcome noise over long beam paths. Second, schemes utilizing X-ray probes, exemplified by the HIBEF project at European XFEL~\cite{Schlenvoigt2016,ahmadiniaz2025towards,Mosman:2021vua,Karbstein2021,formanek2024signatures}, exploit the shorter wavelength to enhance the effect per photon. Here, the principal technical hurdles become the femtosecond- and micrometer-level synchronization of two independent light sources and the preservation of polarization fidelity through X-ray optics. 
Third, theoretical designs using GeV $\gamma$-rays aim to further boost the signal by maximizing the quantum parameter $\chi_\gamma$~\cite{King2016, Bragin2017,nakamiya2017probing,Lv2024,Wistisen2013}. They typically involve a complex three-beam setup (electron-laser collision for photon generation plus a separate laser for probing), which inherits the major synchronization and alignment challenges of the pump-probe architecture.

Notably, while many of these proposals were statistically viable in principle, their practical implementation has been hindered by cumulative technical hurdles—including femtosecond-level synchronization between independent beams, signal degradation during probe-beam transport, and vulnerability to spatiotemporal jitter and unforeseen systematic noises. This recurring pattern underscores the value of an architecture that intrinsically minimizes such implementation complexities.

In parallel, observational evidence suggestive of vacuum birefringence has emerged from extreme astrophysical environments. 
Measurements of polarized optical emission from isolated neutron stars~\cite{Taverna:2022jgl, mignani2016evidence} show signatures consistent with VB predictions in ultra-strong magnetic fields. Similarly, recent analyses of polarized $\gamma\gamma \to e^+e^-$ processes in ultra-peripheral heavy-ion collisions, such as those reported by the STAR collaboration~\cite{brandenburg2023report}, reveal azimuthal asymmetries interpretable as arising from birefringence-like effects where the same electromagnetic field both produces the probing photons and modifies their propagation.

These observations are significant for two reasons. They not only provide indirect yet compelling motivation for VB in nature but also exemplify a unified scenario where photon generation and strong-field interaction are intrinsically co-located---in stark contrast to the separated-beam paradigm of conventional laboratory setups. 
A controlled laboratory detection of VB would therefore serve a dual purpose: confirming a fundamental QED prediction and establishing a crucial benchmark for interpreting astrophysical and heavy-ion signatures.

\begin{figure}[htbp]
  \centering
  \includegraphics[scale=0.4]{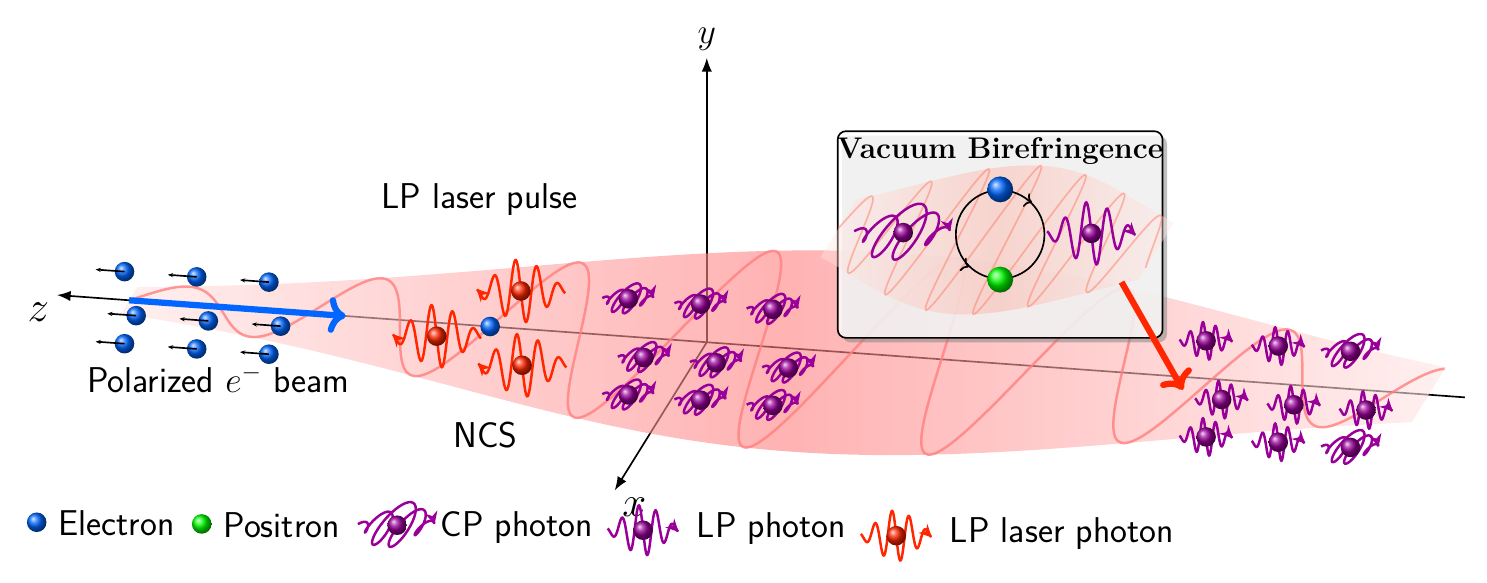}
  \caption{\label{fig:setup} 
   Schematic of the simulation setup. A longitudinally polarized electron beam collides head-on with a high-intensity, linearly polarized laser pulse. 
  The interaction generates high-energy $\gamma$-photons that are initially predominantly circularly polarized (large Stokes parameter $S_2$). 
  As these photons propagate through the remaining laser field, vacuum birefringence (VB) induces a phase difference between polarization components parallel and perpendicular to the laser's electric field. 
  This causes the photons' polarization state to evolve from circular toward elliptical, manifesting as a decrease in $S_2$ and the appearance of a measurable linear polarization component ($S_1$). 
  The evolution of the Stokes parameters $(S_1, S_2, S_3)$ fully captures this polarization transformation.}
\end{figure}

To bridge this gap, we propose a compact, integrated scheme that adopts the unified philosophy observed in nature while operating under well-controlled laboratory conditions. As illustrated in Figure.~\ref{fig:setup}, a relativistic electron beam collides head-on with an ultrastrong laser pulse. Crucially, the \textit{same} laser field serves a dual role: it generates multi-GeV, circularly polarized $\gamma$-photons via nonlinear Compton scattering and simultaneously provides the birefringent medium through which these photons propagate. This ``self-probing'' architecture inherently guarantees perfect spatiotemporal overlap and eliminates the synchronization and beam-transport complexities that challenge conventional pump--probe experiments.

In this Letter, we demonstrate through comprehensive Monte Carlo simulations that this self-probing scheme induces a clear VB signature. Using a longitudinally polarized, 3~GeV electron beam colliding with a petawatt-class laser pulse ($a_0=125$), we observe an average refractive index difference of $\Delta n \approx 1.829 \times 10^{-4}$ for selected photons, accumulating a phase delay of $\Delta\phi \approx 2.343 \times 10^{-2}$~rad over micron-scale paths. This large phase shift, enhanced by the $\gamma$-photon's short wavelength, offers a promising route for the first laboratory observation of VB, with a signal enhanced by nearly 10 orders of magnitude compared to traditional optical experiments.

\section{Simulation Method and Setup}
\label{sec:method}

Our scheme operates in the non-perturbative regime where the quantum nonlinearity parameter for photons, $\chi_\gamma$, approaches unity ($\chi_\gamma \sim 1$). For a high-energy photon in a head-on collision with an intense laser field, $\chi_\gamma$ is given by
\begin{equation}
    \chi_\gamma \approx 0.5741 \frac{\omega_\gamma}{\mathrm{GeV}} \sqrt{\frac{I}{10^{22}\,\mathrm{W/cm}^2}},
    \label{eq:chi_gamma}
\end{equation}
where $\omega_{\gamma}$ is the photon energy and $I$ is the laser intensity. This parameter is boosted to $\chi_\gamma \sim 1$ by the counter-propagating geometry and the GeV-scale photon energy, placing the interaction firmly in the regime where non-perturbative strong-field QED effects dominate. In this regime, we adopt the theoretical framework of Bragin et al.~\cite{Bragin2017}, where the vacuum's refractive and absorptive properties are treated via the full polarization operator. This formalism maps quantum vacuum effects onto an evolution of the photon's Stokes vector: VB manifests as a phase-induced conversion between linear and circular polarization states.

\subsection{Numerical Framework and Implementation}
\label{subsec:numerical_framework}

The simulations are performed using a Monte Carlo method that incorporates all polarization effects \cite{Li2020, Li2019, Zhuang2023}, including vacuum birefringence (VB) and vacuum dichroism (VD), based on the local constant field approximation (LCFA)~\cite{,ritus1985quantum,podszusHighenergyBehaviorStrongfield2019,baier1998electromagnetic,DiPiazza:2018bfu,wu2025}. Quantum probabilities, derived via the Baier–Katkov QED operator method within the LCFA, are employed throughout.

The electron trajectories are computed using a fourth‑order Runge–Kutta method, ensuring accurate propagation in the ultrastrong laser field. The laser field is modeled as a focused Gaussian pulse with a nonparaxial profile that includes corrections up to order $(w_0/z_r)^3$, where $w_0$ is the beam waist and $z_r$ the Rayleigh length~\cite{Salamin2002}.

At each time step, the electron (or positron) Stokes polarization vector $\bm{\zeta}_{\mathrm{p}}$ evolves into $\bm{\zeta}_{\mathrm{f}}^{\mathrm{R}}$ after photon emission. Here $\bm{\zeta}_{\mathrm{p}}$ ($\bm{\zeta}_{\mathrm{f}}^{\mathrm{R}}$) denotes the electron spin polarization vector in a mixed state before (after) radiation, expressed in the electron rest frame. If no photon is emitted during the step, the electron spin evolves into $\bm{\zeta}_{\mathrm{f}}^{\mathrm{NR}}$ according to the no‑emission probability, which originates from the one‑loop contribution to the electron mass operator.

Spin precession between emission events is governed by the Thomas–Bargmann–Michel–Telegdi (TBMT) equation, including the field‑dependent anomalous magnetic moment from one‑loop vertex corrections\cite{thomas1926motion,thomas1927kinematics,bargmann1959precession,walser2002spin}. Between quantum events, the electron (or positron) dynamics in the laser field are described classically by the Lorentz equation. The Stokes vector of a newly created photon or electron–positron pair is determined by the polarization‑resolved photon‑emission or pair‑production probability.

\subsubsection*{Photon polarization at emission}

The polarization state of a newly emitted photon is obtained from the polarization‑resolved probability of nonlinear Compton scattering. Within the LCFA, the differential emission probability reads \cite{Li2020}
\begin{equation}
\frac{d^{2}W_{fi}}{du\,d\eta}= \frac{W_{R}}{2}\bigl(F_{0}+ S_{1}F_{1}+ S_{2}F_{2}+ S_{3}F_{3}\bigr),
\label{eq:pol_emission_prob}
\end{equation}
where 
\[
W_{R}= \frac{\alpha m}{8\sqrt{3}\pi\lambda_{c}\,(k\cdot p_{i})(1+u)^{3}},\qquad 
u=\frac{\omega_{\gamma}}{\epsilon_{i}-\omega_{\gamma}} .
\]
Here $\epsilon_{i}$ is the electron energy before radiation, $\alpha = 1/137$ the fine‑structure constant, $\lambda_{c}$ the Compton wavelength, $\eta=k\cdot r$ the laser phase, and $p_{i}$, $k$, $r$ are the four‑vectors of the electron momentum before radiation, the laser wave vector, and the coordinate, respectively. The photon Stokes parameters $S_{1},S_{2},S_{3}$ follow the convention of Ref.~\cite{Bragin2017}: $S_2$ denotes circular polarization (helicity), whereas $S_1$ and $S_3$ represent linear polarization with $S_3$ defined along the laser’s $x$-axis.

The mean Stokes parameters of the emitted photon are given by
\begin{equation}
S_{1}=\frac{F_{1}}{F_{0}},\qquad 
S_{2}=\frac{F_{2}}{F_{0}},\qquad 
S_{3}=\frac{F_{3}}{F_{0}} .
\label{eq:mean_stokes}
\end{equation}
The functions $F_{0},F_{1},F_{2},F_{3}$ are obtained after summing over final electron polarizations:
\begin{align}
F_{0}&= \frac{\alpha m^{2}}{\sqrt{3}\,\pi\epsilon_i}\,
\frac{du}{(1+u)^{3}}
\Bigl\{-(1+u)\int_{\frac{2u}{3\chi_e}}^{\infty}\!dx\,K_{1/3}(x)
      +2\Bigl(1+u \nonumber\\
       &\qquad +\frac{u^{2}}{2}\Bigr)K_{2/3}\!\Bigl(\frac{2u}{3\chi_e}\Bigr) -u\Bigl(\bm{\zeta}_{\mathrm{i}}\cdot\hat{\bm e}_{\mathrm{2}}\Bigr)
            K_{1/3}\!\Bigl(\frac{2u}{3\chi_e}\Bigr)\Bigr\}, \label{eq:F0_full}\\[4pt]
F_{1}&= \frac{\alpha m^{2}}{\sqrt{3}\,\pi\epsilon_i}\,
\frac{du}{(1+u)^{3}}
\Bigl\{(1+u)K_{2/3}\!\Bigl(\frac{2u}{3\chi_e}\Bigr)
      -u(1\nonumber\\
&\qquad+u)\Bigl(\bm{\zeta}_{\mathrm{i}}\cdot\hat{\bm e}_{\mathrm{2}}\Bigr)
            K_{1/3}\!\Bigl(\frac{2u}{3\chi_e}\Bigr)\Bigr\}, \label{eq:F1_full}\\[4pt]
F_{2}&=-\frac{\alpha m^{2}}{\sqrt{3}\,\pi\epsilon_i}\,
\frac{du}{(1+u)^{3}}
\Bigl\{\bigl(\bm{\zeta}_{\mathrm{i}}\cdot\hat{\bm e}_{\mathrm{v}}\bigr)
       \int_{\frac{2u}{3\chi_e}}^{\infty}\!dx\,K_{1/3}(x)
      -(2\nonumber\\
      &\qquad+u)u\,K_{2/3}\!\Bigl(\frac{2u}{3\chi_e}\Bigr)\Bigr\}, \label{eq:F2_full}\\[4pt]
F_{3}&= \frac{\alpha m^{2}}{\sqrt{3}\,\pi\epsilon_i}\,
\frac{du}{(1+u)^{3}}
\Bigl\{(1+u)K_{2/3}\!\Bigl(\frac{2u}{3\chi_e}\Bigr)
      -u(1\nonumber\\
      &\qquad+u)\Bigl(\bm{\zeta}_{\mathrm{i}}\cdot\hat{\bm e}_{\mathrm{2}}\Bigr)
            K_{1/3}\!\Bigl(\frac{2u}{3\chi_e}\Bigr)\Bigr\}, \label{eq:F3_full}
\end{align}
where $\hat{\bm e}_{\mathrm{v}}$ is the unit vector along the electron velocity, $\hat{\bm e}_{1}$ the unit vector along the electron acceleration (which, for our counter‑propagating geometry with laser polarization along $x$, aligns with the $x$-axis), and $\hat{\bm e}_{2}=\hat{\bm e}_{\mathrm{v}}\times\hat{\bm e}_{1}$. In the expressions above, $\chi_e$ is the electron quantum strong‑field parameter, $\epsilon_i$ the electron energy before emission, and $K_{\nu}$ denotes the modified Bessel function of the second kind of order $\nu$.

In the present work the electron beam is initially fully longitudinally polarized, i.e. $\bm{\zeta}_{\mathrm{i}}=(0,0,1)$. For this configuration the velocity is along the $-z$‑axis and the acceleration (hence the major axis of the polarization ellipse) is along the $x$‑axis, leading to
\[
\bm{\zeta}_{\mathrm{i}}\cdot\hat{\bm e}_{\mathrm{v}}=-1,\qquad
\bm{\zeta}_{\mathrm{i}}\cdot\hat{\bm e}_{2}=0 .
\]
Consequently the general expressions simplify and the initial Stokes parameters of the emitted photon, denoted as $\mathbf{S}^i = (S^i_1, S^i_2, S^i_3)$, become
\begin{align}
S^i_{1}&= 0, \label{eq:S1_long} \\[6pt]
S^i_{2}&= \frac{\displaystyle \int_{\frac{2u}{3\chi_e}}^{\infty} dx\, K_{1/3}(x) + (2+u)u K_{2/3}\!\left(\frac{2u}{3\chi_e}\right)}
            {\displaystyle -(1+u)\int_{\frac{2u}{3\chi_e}}^{\infty} dx\, K_{1/3}(x)
             +2\bigl(1+u+\frac{u^{2}}{2}\bigr)K_{2/3}\!\bigl(\frac{2u}{3\chi_e}\bigr)}, \label{eq:S2_long} \\[6pt]
S^i_{3}&= \frac{(1+u)\,K_{2/3}\!\bigl(\frac{2u}{3\chi_e}\bigr)}
            {\displaystyle -(1+u)\int_{\frac{2u}{3\chi_e}}^{\infty}\!dx\,K_{1/3}(x)
             +2\bigl(1+u+\frac{u^{2}}{2}\bigr)K_{2/3}\!\bigl(\frac{2u}{3\chi_e}\bigr)}. \label{eq:S3_long}
\end{align}

\subsection{Polarization Evolution Implementation}
\label{subsec:polarization_evolution}

As photons propagate through the ultrastrong laser field, their polarization state evolves due to VP effects. For a photon with an initial Stokes vector $\mathbf{S}^i = (S^i_1, S^i_2, S^i_3)$, the VB and VD modify its polarization during propagation.

\subsubsection*{Vacuum birefringence}
The VB effect is implemented within the LCFA framework following the formulation of Bragin et al.~\cite{Bragin2017}. The VB‑induced rotation in the $(S_1, S_2)$ plane yields the final Stokes components
\begin{equation}
    \begin{pmatrix} S^f_{1} \\ S^f_{2} \end{pmatrix}
    = \begin{pmatrix} \cos\delta\phi & -\sin\delta\phi \\ \sin\delta\phi & \cos\delta\phi \end{pmatrix}
      \begin{pmatrix} S^i_{1} \\ S^i_{2} \end{pmatrix},
    \label{eq:VB_rotation}
\end{equation}
while the $S_3$ component remains unchanged under VB. Here $\delta\phi = \phi_1 - \phi_2$ is the accumulated phase difference between the two polarization eigenmodes, computed from
\begin{equation}
    \phi_i = -\frac{1}{2kq}\mathrm{Re}\left[p_i(\chi_{\gamma})\right]\Delta\eta,
\end{equation}
with
\begin{equation}
    \begin{bmatrix} p_1(\chi_{\gamma}) \\ p_2(\chi_{\gamma}) \end{bmatrix}
    = \frac{\alpha m^2}{3\pi} \int_{-1}^{+1} d\nu
      \begin{bmatrix} w-1 \\ w+2 \end{bmatrix}
      \frac{f'(u)}{u},
\end{equation}
$w = 4/(1-\nu^2)$, $u = [w/\chi_{\gamma}]^{2/3}$, $f(u) = \pi[\mathrm{Gi}(u) + i\mathrm{Ai}(u)]$, and $\eta = \omega_0 t - k_0 z$ being the laser phase.

\subsubsection*{Vacuum dichroism}
The VD effect, which corresponds to polarization‑dependent attenuation of photons via the imaginary part of the VP loop, is implemented through a no‑pair‑production evolution as described in Ref.~\cite{Zhuang2023}. 
At each time step, for a photon with current Stokes vector $\mathbf{S}^i$, the loop probability is given by
\begin{equation}
    W_{\mathrm{loop}} = F_0^{\mathrm{L}} + \mathbf{S}^i \cdot \mathbf{F}^{\mathrm{L}},
\end{equation}
where
\begin{align}
    F_0^{\mathrm{L}} &= \frac{1}{2}\Bigg\{1 - \int_0^1 dx\,C_{\mathrm{p}}\Bigg[\Bigg(\frac{x}{1-x} + \frac{1-x}{x}\Bigg)K_{2/3}(\rho)  \nonumber\\
       &\qquad + \int_{\rho}^{\infty}\!dx\,K_{1/3}(x) - S^i_3 K_{2/3}(\rho)\Bigg]\Delta t\Bigg\}, \\[4pt]
    \mathbf{F}^{\mathrm{L}} &= \frac{1}{2}\mathbf{S}^i\Bigg\{\mathbf{S}^i\Bigg[1 - \int_0^1 dx\,C_{\mathrm{p}}\Bigg(\int_{\rho}^{\infty}\!dx\,K_{1/3}(x) + \Bigg(\frac{x}{1-x}  \nonumber\\
       &\qquad + \frac{1-x}{x}\Bigg)K_{2/3}(\rho)\Bigg)\Delta t\Bigg]\Bigg\}.
\end{align}
Here $C_{\mathrm{p}} = \alpha m^{2} / (\sqrt{3}\pi \omega_{\gamma})$, $\rho = 2/[3\chi_{\gamma} x(1-x)]$, and $K_{2/3}$, $K_{1/3}$ are the modified Bessel functions of the second kind. The Stokes vector after the time step, $\mathbf{S}^f = (S^f_1, S^f_2, S^f_3)$, is then obtained as
\begin{equation}
    \mathbf{S}^f = \frac{\mathbf{F}^{\mathrm{L}}}{F_0^{\mathrm{L}}}.
    \label{eq:VD_evolution}
\end{equation}
This treatment accounts for the differential absorption of photons depending on their polarization states, which ultimately leads to a net polarization enhancement.
To simplify the equation, we can also write the evolution of the Stokes parameters in a differential form as follows\cite{dai2024fermionic}:
\begin{align}
\frac{dS_1}{dt} &= - \int \frac{\alpha m^2 d\varepsilon}{\sqrt{3}\pi\omega_{\gamma}^2} S_3 S_1 K_{\frac{2}{3}}(\rho), \\
\frac{dS_2}{dt} &= - \int \frac{\alpha m^2 d\varepsilon}{\sqrt{3}\pi\omega_{\gamma}^2} S_3 S_2 K_{\frac{2}{3}}(\rho), \\
\frac{dS_3}{dt} &= \int \frac{\alpha m^2 d\varepsilon}{\sqrt{3}\pi\omega_{\gamma}^2} (1 - S_3^2) K_{\frac{2}{3}}(\rho),
\label{eq:VD}
\end{align}

\subsection{Simulation Parameters (Baseline)}
\label{subsec:parameters}

Unless stated otherwise, all simulations presented in this work employ the following baseline parameters.

The electron beam propagates along $-z$ and contains $N_e = 10^6$ macro‑particles. It has a cylindrical profile with a longitudinal length $L_e = 3\,\mu\mathrm{m}$ and transverse Gaussian widths $\sigma_{x,y} = 0.3\,\mu\mathrm{m}$. The initial mean energy is $\epsilon_i = 3\,\mathrm{GeV}$, with a $10\%$ (FWHM) energy spread and a $3\,\mathrm{mrad}$ (FWHM) angular divergence. The beam is initially fully longitudinally polarized: $\bm{\zeta}_{\mathrm{i}} = (0,0,1)$.
These parameters are well within the capabilities of current laser-wakefield accelerators~\cite{Gonsalves2019}.

The laser pulse propagates along $+z$ and is linearly polarized along the $x$-axis. Its parameters are: wavelength $\lambda_0 = 1\,\mu\mathrm{m}$, focal spot size $w_0 = 5\,\mu\mathrm{m}$, and duration $\tau = 20\,T_0$, where $T_0 = 2\pi/\omega_0$ is the optical period. The peak intensity reaches $I_0 \approx 2.16 \times 10^{22}\,\mathrm{W}/\mathrm{cm}^2$, corresponding to a normalized vector potential $a_0 = eE_0/(m_e c \omega_0) = 125$.
The laser parameters are chosen to be compatible with the capabilities of near-future petawatt-class laser systems~\cite{tanaka2020current}.
For this baseline configuration, the electron quantum strong‑field parameter is
\[
\chi_e = \frac{2\gamma\,a_0\,\hbar\omega_0}{m_e c^2} \approx 3.6,
\]
where $\gamma = \epsilon_i/(m_e c^2) \approx 5870$ is the electron Lorentz factor. This places the interaction well into the non‑perturbative QED regime ($\chi_e \gtrsim 1$). For a representative high‑energy photon of $\omega_\gamma = 1\,\mathrm{GeV}$, the corresponding photon quantum parameter is
\[
\chi_\gamma = \frac{\omega_\gamma}{m_e c^2}\,\frac{2a_0\,\hbar\omega_0}{m_e c^2} \approx 1.2,
\]
indicating that both electrons and high‑energy photons experience significant strong‑field QED effects. Any deviations from this baseline parameter set are explicitly noted in the respective sections.

\section{Simulation Results}
\label{sec:results}

\subsection{Angular Selection and Polarization Signatures}
\label{subsec:angular_selection}

\begin{figure}[htbp]
  \centering
  \begin{overpic}[width=0.23\textwidth]{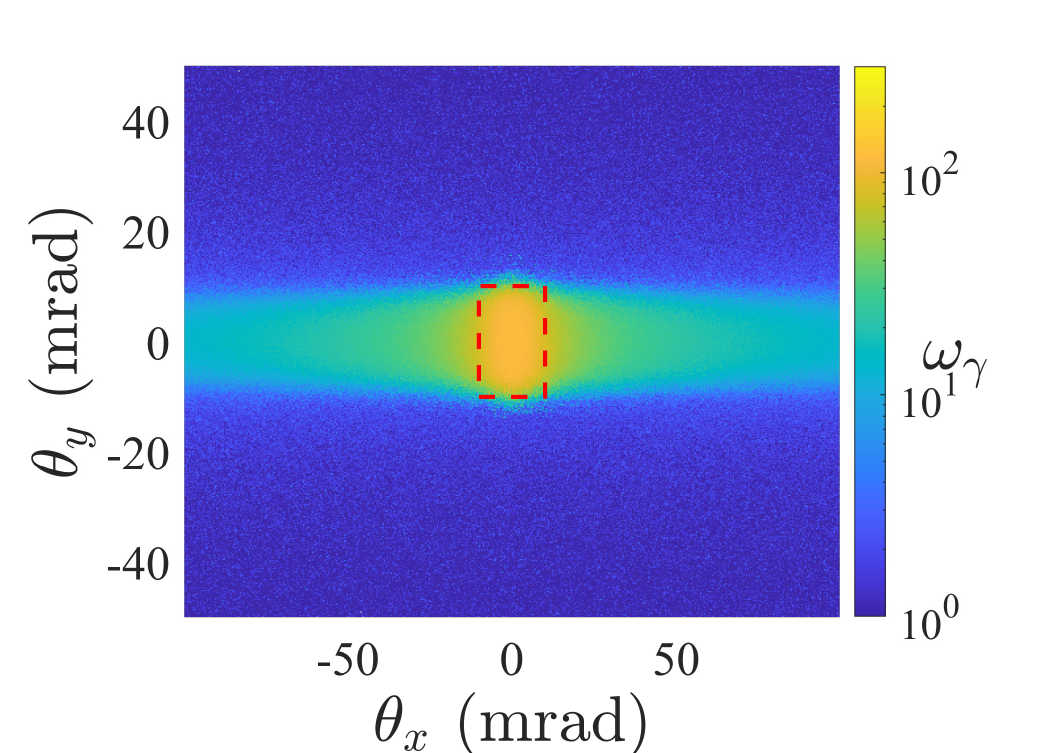}
      \put(5, 70){\small (a)}
      \label{fig:sub1}
  \end{overpic}
  \hspace{2mm}
  \begin{overpic}[width=0.23\textwidth]{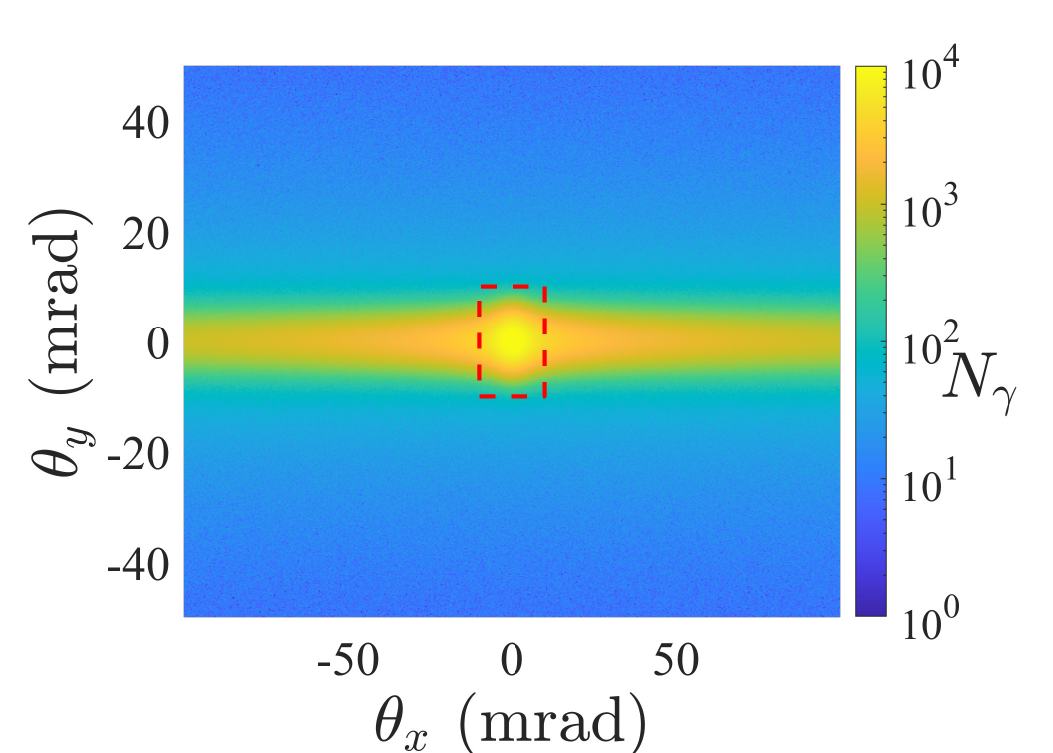}
      \put(5, 70){\small (b)}
      \label{fig:sub2}
  \end{overpic}
  \hspace{2mm}
  \begin{overpic}[width=0.23\textwidth]{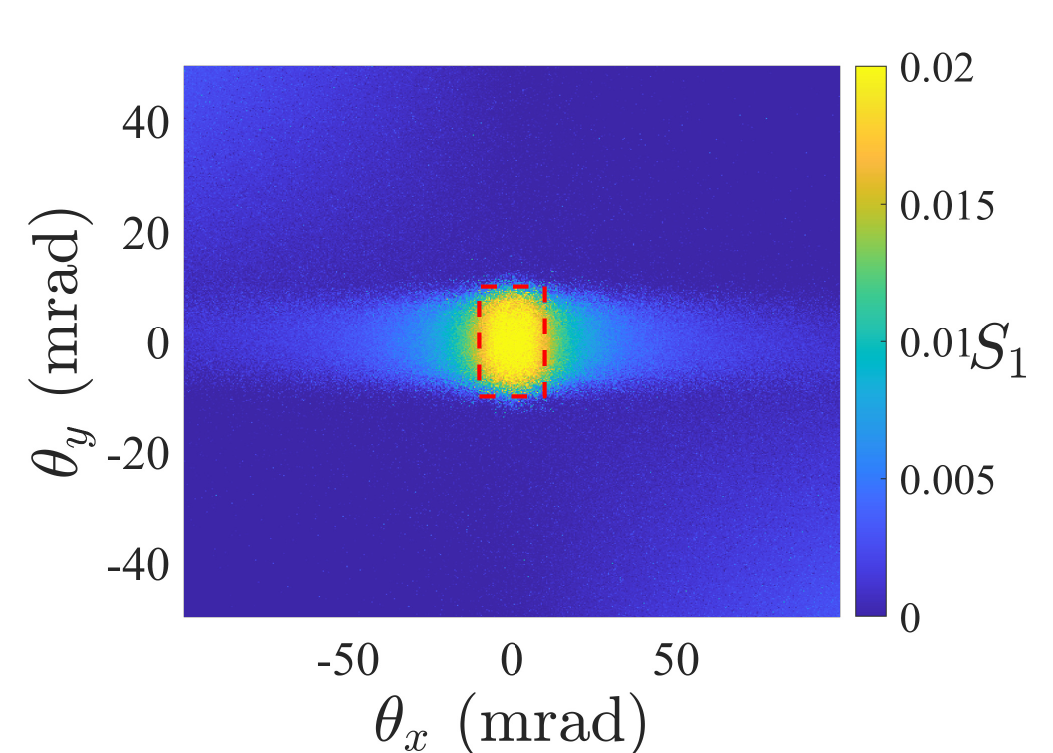}
      \put(5, 70){\small (c)}
      \label{fig:sub3}
  \end{overpic}
  \hspace{2mm}
  \begin{overpic}[width=0.23\textwidth]{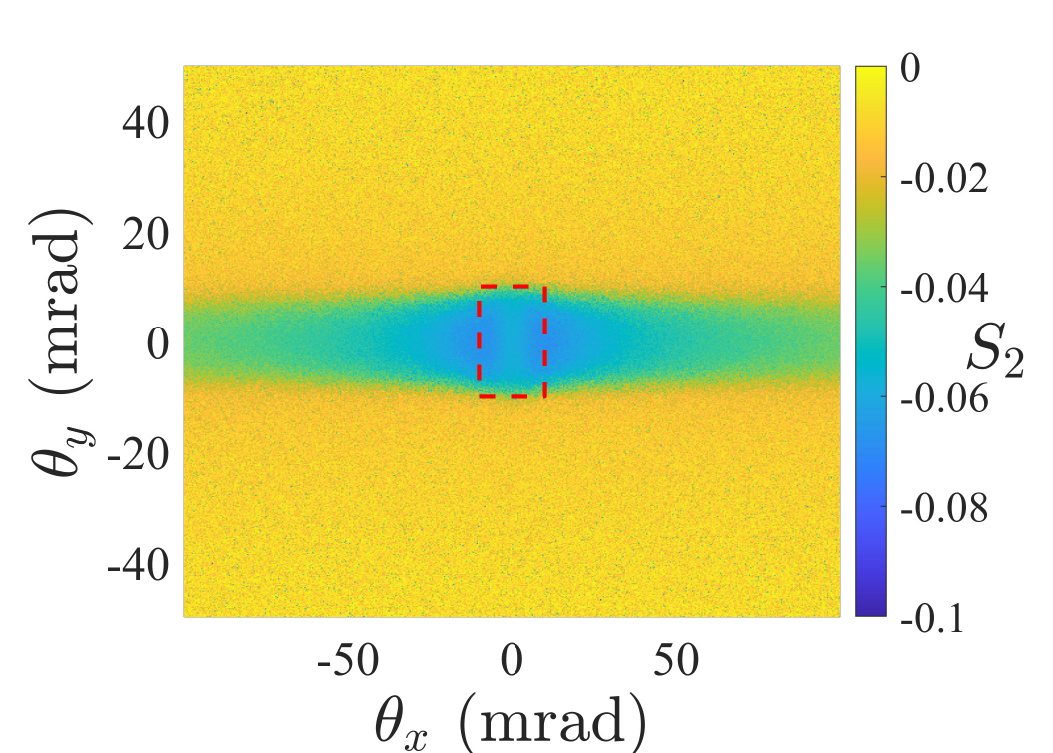}
      \put(5, 70){\small (d)}
      \label{fig:sub4}
  \end{overpic}
  
  \vspace{2mm}
  
  \begin{overpic}[width=0.23\textwidth]{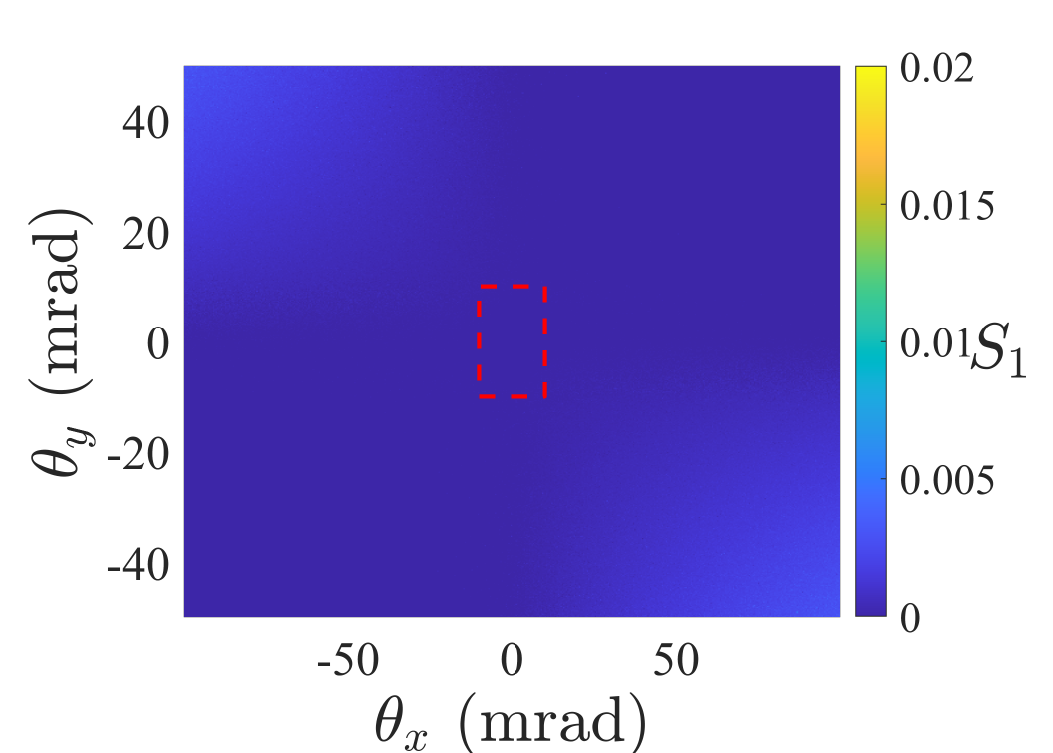}
      \put(5, 70){\small (e)}
      \label{fig:sub5}
  \end{overpic}
  \hspace{2mm}
  \begin{overpic}[width=0.23\textwidth]{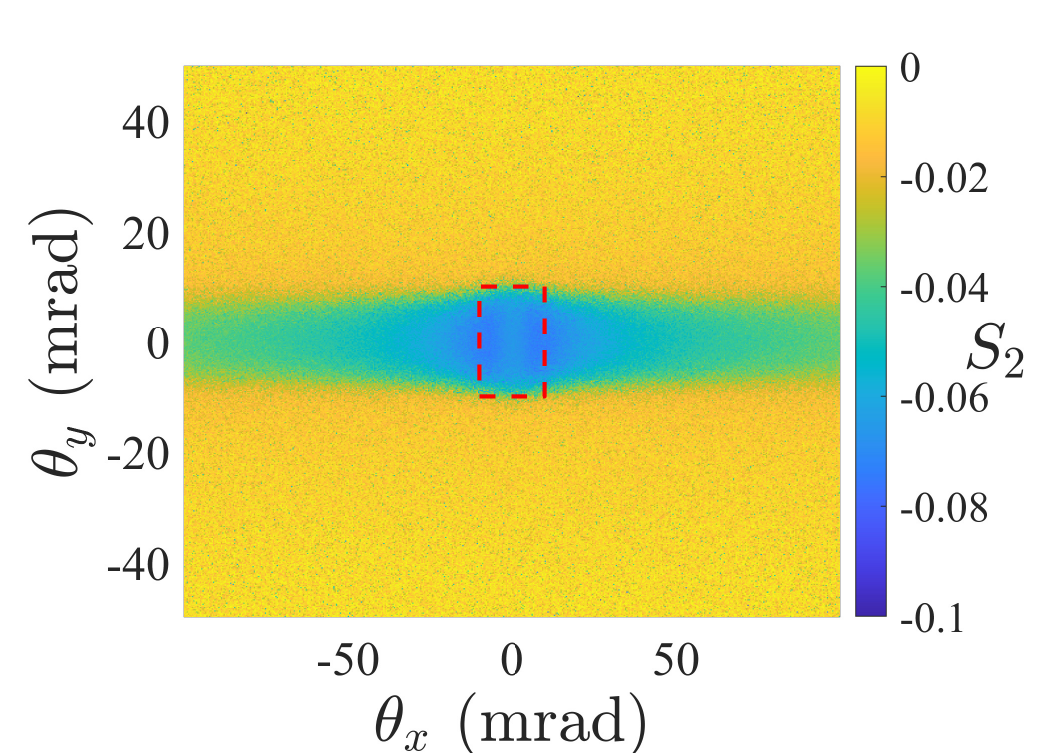}
      \put(5, 70){\small (f)}
      \label{fig:sub6}
  \end{overpic}
  \caption{Angular distributions of (a) photon energy, (b) photon number, (c) Stokes parameter $S_1$, and (d) $S_2$ from the full simulation including VP. Panels (e) and (f) show the corresponding $S_1$ and $S_2$ distributions when VP effects are artificially disabled. The dashed rectangle in all panels indicates the selection region applied to isolate high-energy photons for analysis, spanning $\pm10$~mrad in both horizontal and vertical divergence angles ($\theta_x, \theta_y \in [-10,10]$~mrad).}
  \label{fig:angular_distribution}
\end{figure}

Figure~\ref{fig:angular_distribution} presents the angular distributions of key photon properties obtained from our Monte Carlo simulation. A clear correlation between photon energy, emission angle, and polarization state is evident, which we exploit to isolate the high-energy photon population that carries the strongest VB signal.

Photons generated via nonlinear Compton scattering exhibit a strong forward peaking. As shown in Figure.~\ref{fig:angular_distribution}(a) and (b), both the mean photon energy and the photon flux are concentrated within a narrow cone of $\sim 10$~mrad half-angle. This behavior arises from two combined effects: first, electrons undergoing radiative energy loss emit higher-energy photons early in the interaction, and these photons are scattered into smaller angles due to the large Lorentz factor of the beam ($\gamma_e \approx 6000$); second, in a linearly polarized laser field, the electron's transverse velocity is phase-shifted by $\pi/2$ relative to the electric field, resulting in negligible net transverse momentum and confining the radiation predominantly along the instantaneous electron velocity direction. Consequently, the central angular region contains both the highest photon energies and the majority of the photon yield.

Motivated by the fact that VB scales linearly with photon energy (via $\chi_\gamma \propto \omega_\gamma$), we restrict our analysis to photons within the rectangular region of $\pm10$~mrad in both transverse divergence angles (marked in all panels of Fig.~\ref{fig:angular_distribution}). This selection retains over $85\%$ of photons with energy above $1$~GeV while discarding more than $90\%$ of the low-energy background ($\sim 10$–$100$~MeV). Within this region, the average photon energy is $107$~MeV, and it contains $1.51\times10^7$ photons, of which $1\times10^5$ exceed $1$~GeV. Experimentally, such a narrow angular cut can be conveniently implemented with a collimator placed on the beam axis, making this region both physically and diagnostically favorable.

The decisive signature of VB is revealed by comparing the angular distributions of the Stokes parameters $S_1$ and $S_2$ obtained with and without VP. In the absence of VP [Figure.~\ref{fig:angular_distribution}(e) and (f)], the emitted photons are almost purely circularly polarized: $S_1$ remains negligible (average $0.00001$) while $S_2$ exhibits a pronounced negative value (average $-0.069$), consistent with the helicity transfer from the longitudinally polarized electron beam~\cite{Li2020}. When VP is included [Figure.~\ref{fig:angular_distribution}(c) and (d)], a non-zero $S_1$ component emerges across the central angular region (average $0.019$), directly reflecting the conversion of circular into linear polarization induced by the birefringent vacuum. The accompanying $S_2$ magnitude shows only a slight modification (average from $-0.069$ to $-0.061$), indicating that the VB-induced phase shift is still modest at the mean photon energy of the selected sample.

Although the average $S_1$ within the full selected region is only $0.019$, this value is strongly weighted by the numerous sub-GeV photons, for which $\chi_\gamma$ is small and VB is weak. 
As we demonstrate in the next subsection, the substantial photon yield within the selected region effectively compensates for the modest average $S_1$ signal, providing a robust statistical foundation for the detection of vacuum birefringence.

\subsection{Energy-Resolved VB Signature}
\label{subsec:energy_resolved}

\begin{figure}[htbp] 
  \centering
  \begin{overpic}[width=0.23\textwidth]{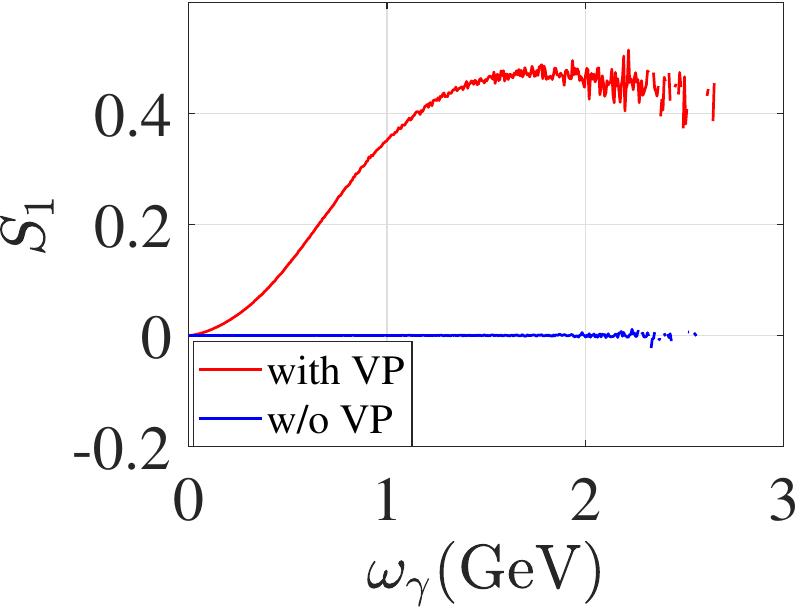} 
      \put(7, 75){\large (a)}
  \end{overpic}
  \begin{overpic}[width=0.23\textwidth]{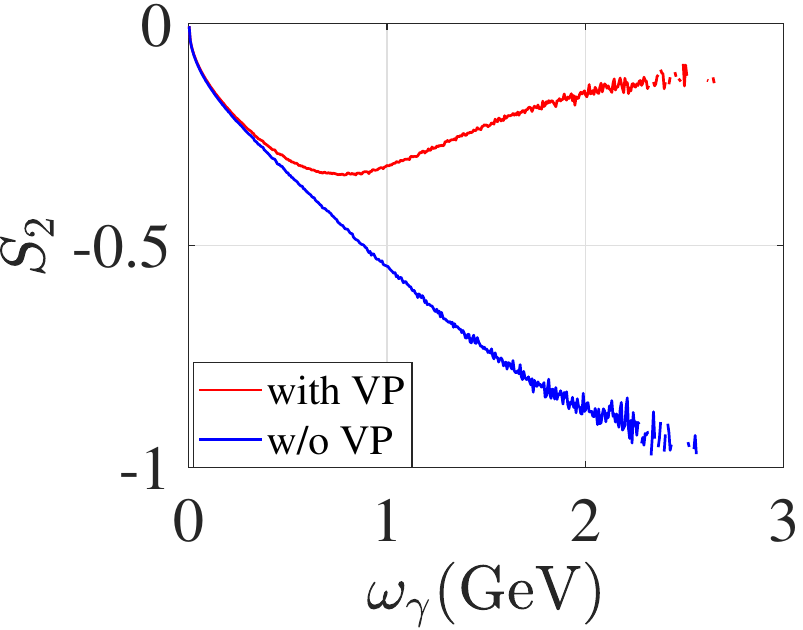}
      \put(2, 75){\large (b)}
  \end{overpic}
  \hspace{3mm}
  \begin{overpic}[width=0.23\textwidth]{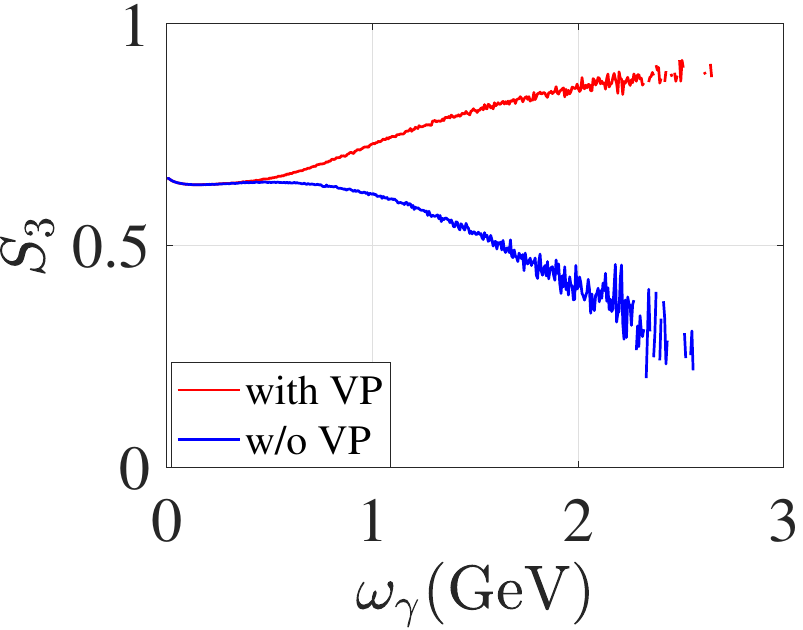}
    \put(3, 75){\large (c)}
  \end{overpic}
  \begin{overpic}[width=0.23\textwidth]{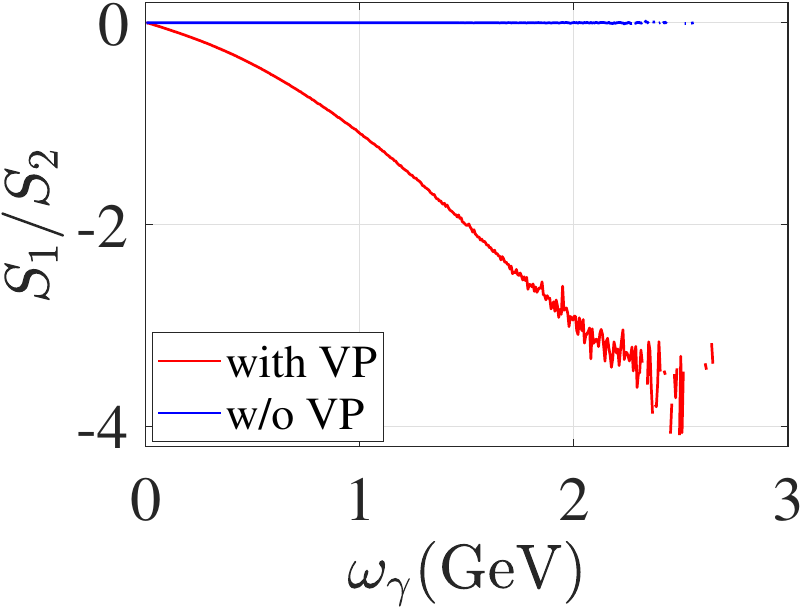}
    \put(0, 75){\large (d)}
  \end{overpic}
  \caption{Signature of VB. Stokes parameters (a) $S_1$, (b) $S_2$, (c) $S_3$, and (d) $S_1/S_2$ as functions of photon energy for photons within the angular selection region of Figure.~\ref{fig:angular_distribution}. Red curves: results including VP (both VB and VD). Blue curves: results with VP artificially disabled. }
  \label{fig:vb_signature}
\end{figure}

To quantitatively assess the VB effect and its dependence on photon energy, we further analyse the selected photons by binning them in energy. Figure~\ref{fig:vb_signature} plots the Stokes parameters $S_1$, $S_2$, $S_3$, and the ratio $S_1/S_2$ as functions of photon energy $\omega_\gamma$, comparing simulations with and without VP.

When VP is disabled (blue curves), $S_1$ remains essentially zero across the entire energy range, while $|S_2|$ approaches unity at high energies, see Figure.~\ref{fig:vb_signature} (a) and (b), indicating strong circular polarization inherited from the longitudinally polarized electron beam. 
In stark contrast, when VP is active (red curves), a pronounced $S_1$ component appears for photons above $\omega_\gamma \gtrsim 0.5$~GeV, rising to $\sim 0.45$ at $2$~GeV. This $S_1$ generation is accompanied by a substantial suppression of $|S_2|$, which drops from $\approx -0.9$ to $\approx -0.15$ over the same energy range. The simultaneous $S_2 \rightarrow S_1$ conversion and the reduction of circular polarization are the unambiguous hallmarks of VB.

The non-zero $S_3$ component visible in Figure.~\ref{fig:vb_signature}(c) (blue curve) is a well-known feature of nonlinear Compton scattering in a linearly polarized laser field~\cite{Li2020}: it arises from the polarization-dependent emission probability and does not require vacuum effects. Additionally, Figure.~\ref{fig:vb_signature}(c) reveals a significant increase in $S_3$ at multi-GeV energies when VP is included. This effect arises from VD \cite{wu2025,li2020production,ivanov2005complete}: the polarization-dependent attenuation of photons due to nonlinear Breit–Wheeler pair production, which preferentially absorbs one helicity component and thereby modifies the ensemble-averaged Stokes vector.

The ratio $S_1/S_2$, shown in Figure.~\ref{fig:vb_signature}(d), further emphasizes the growing influence of VB with photon energy. While the ratio remains near zero without VP, it reaches $\sim -3$ at $2$~GeV when VP is included, directly quantifying the relative strength of the induced linear polarization compared to the residual circular component.

These energy-resolved results confirm that the VB signal is most pronounced for GeV-scale photons. Although such high-energy photons are less abundant than their sub-GeV counterparts, their large per-photon $S_1$ yield makes them the optimal carriers of the VB signature. The trade-off between signal strength and photon statistics is systematically addressed in the parameter optimization discussed in Sec.~\ref{sec:optimization}.

\begin{figure}[htbp]
  \centering
  \begin{overpic}[width=0.225\textwidth]{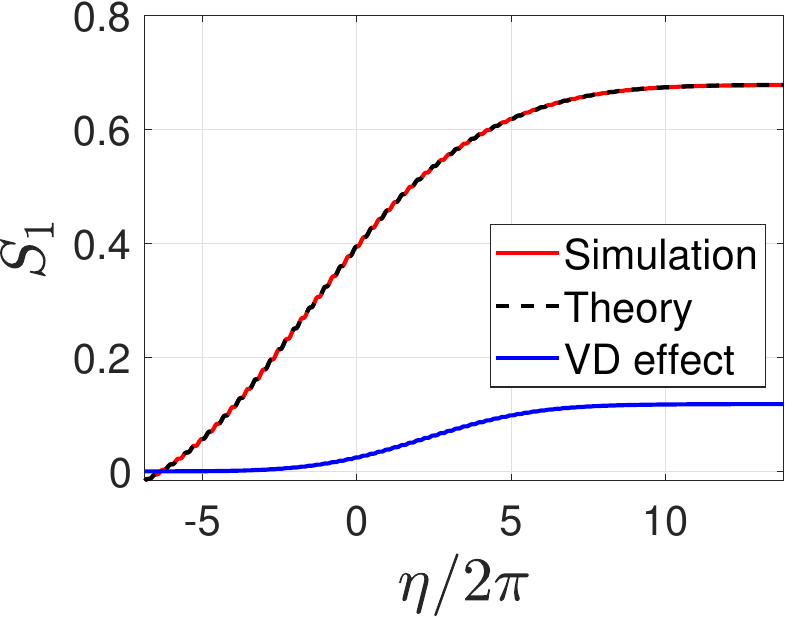}
      \put(0, 75){\small (a)}
      \label{fig:t_polar}
  \end{overpic}
  \hspace{2mm}
  \begin{overpic}[width=0.235\textwidth]{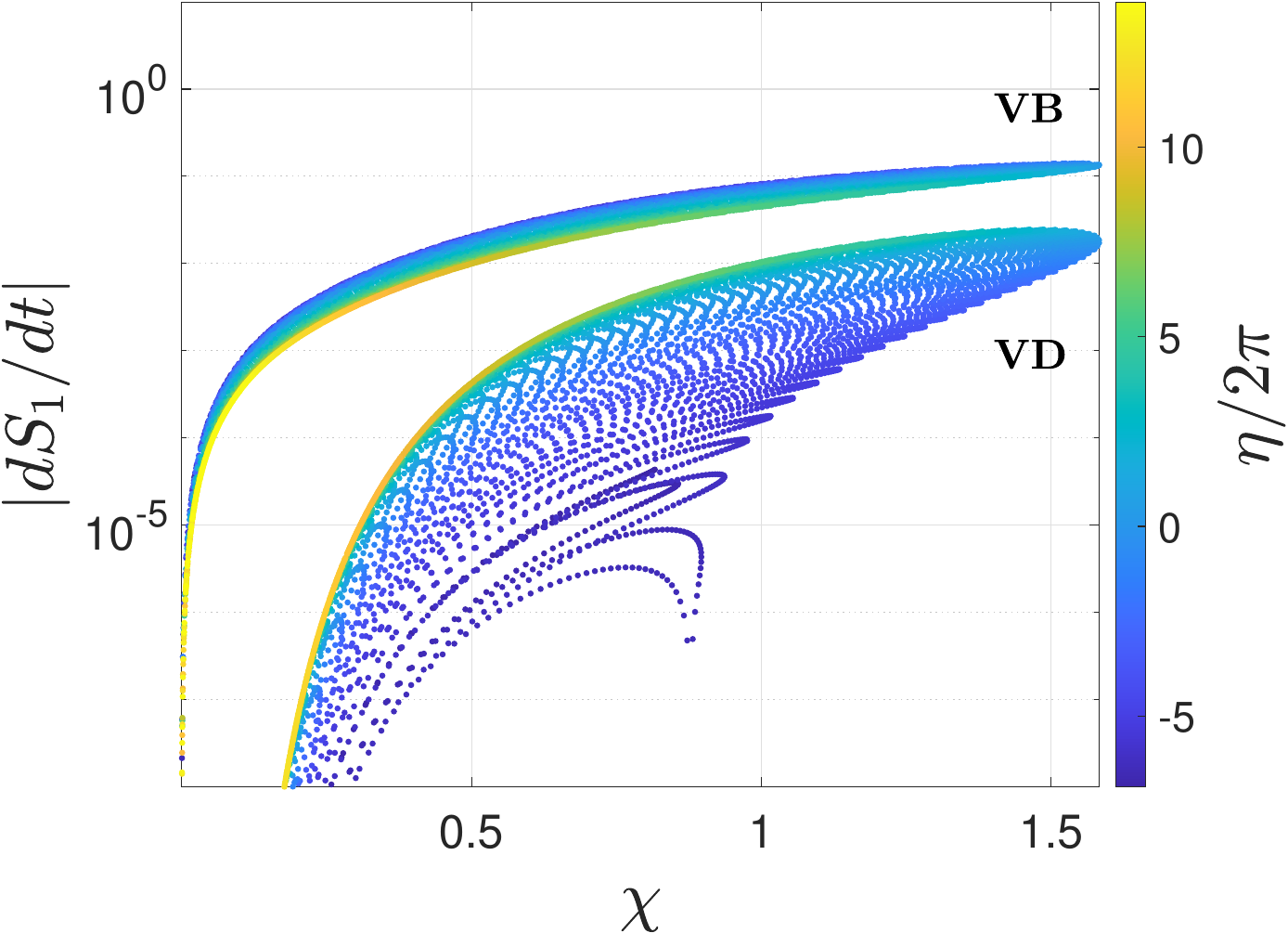}
      \put(0, 75){\small (b)}
      \label{fig:ds1dt_vs_chi}
  \end{overpic}
\caption{(a) Evolution of the Stokes parameter $S_1$ as a function of the laser phase $\eta = \omega_0 t - k_0 z $ for a 1~GeV photon. Red solid line: full Monte Carlo simulation including both VB and VD. Blue dashed line: deterministic integration of the VP-induced phase shift based on the nonperturbative polarization operator (Eqs.~\ref{eq:VB_rotation} and Eqs.~\ref{eq:VD}). Blue solid line: contribution from VD alone, obtained by Eqs.~\ref{eq:VD}. 
(b) Comparison of the instantaneous polarization rotation rates $dS_1/dt$ arising from VB  and VD  as a function of the quantum parameter $\chi_\gamma$ for the same 1.341 GeV photon. The dots represent simulation data. The color gradient of the points reflects the evolution of the laser phase (interaction time).}
 \label{fig:S1_Evolution}
\end{figure}

Figure~\ref{fig:S1_Evolution}(a) shows the buildup of linear polarization as a high-energy photon ($\omega_\gamma = 1.341$ GeV) traverses the laser pulse. The full simulation (red solid curve) exhibits a monotonic increase of $S_1$, reaching $\approx 0.7$ at the end of the interaction. This evolution is almost perfectly reproduced by the deterministic integration of the phase shift $\delta\phi$ derived from the polarization operator (blue dashed curve), confirming the accuracy of our Monte Carlo implementation. 

In contrast, the contribution arising solely from VD (blue solid curve), computed by artificially disabling the VB-induced rotation, remains below 0.119 throughout the entire interaction. For this specific single-photon track, the integrated change due to VD is $(\Delta S_1, \Delta S_2, \Delta S_3) = (-0.118, 0.078, 0.361)$. 
We emphasize that these values represent the specific dynamical response of a single 1.341 GeV photon rather than an energy-averaged ensemble. Although VD is not negligible for GeV-scale photons, it is clearly subdominant; the vast majority of the $S_1$ signal originates from the coherent phase accumulation of VB.

To further resolve the local interaction dynamics, the instantaneous polarization change rates $dS_1/dt$ arising from  VB and VD are plotted against the quantum parameter $\chi_\gamma$ in Fig.~\ref{fig:S1_Evolution}(b) for the same 1.341 GeV photon track. 
Each point in the scatter plot represents a specific instant along the trajectory, with the color gradient encoding the evolution of the laser phase. 
The VB-induced rotation rate dominates the interaction, exhibiting a characteristic scaling consistent with the $\chi_\gamma^{2/3}$ dependence predicted by the local constant field approximation\cite{DiPiazza:2018bfu,Sevostyanov2021,Aleksandrov2023}. 
In contrast, the change rate due to VD remains significantly lower across the entire range of experienced $\chi_\gamma$, even at the peak of the pulse. 
The clear separation between these two rates throughout the photon's path provides direct dynamical evidence that the observed $S_1$ signal is a primary manifestation of the dispersive vacuum ellipticity.
This differential analysis reinforces the conclusion that the macroscopic polarization signal can be reliably attributed to VB.

\section{Measurement Scheme and Parameter Optimization}
\label{sec:optimization}

To experimentally observe the VB signal, we adopt a polarimetry scheme based on the azimuthal distribution of electron-positron pairs produced when the $\gamma$-photons strike a high-$Z$ converter (e.g., tungsten) \cite{Bragin2017}. The differential cross-section for pair production in the converter is given by:
\begin{equation}
\mathrm{d}\sigma_{\mathrm{pp}}=\frac{\mathrm{d}\varphi}{2\pi}\left\{S_0\sigma_0+[S_1\sin(2\varphi)+S_3\cos(2\varphi)]\sigma_1\right\},
\end{equation}
where $\varphi$ is the azimuthal angle of the produced pairs, $\sigma_0$ and $\sigma_1$ are the unpolarized and polarized cross-section coefficients, and $S_i$ are the Stokes parameters of the photon beam. The linear polarization $S_1$ manifests as an azimuthal modulation, which can be extracted via the asymmetry parameter $R_B$:
\begin{equation}
R_B = \frac{(N_{\pi/4}+N_{5\pi/4})-(N_{3\pi/4}+N_{7\pi/4})}{(N_{\pi/4}+N_{5\pi/4})+(N_{3\pi/4}+N_{7\pi/4})}.
\end{equation}
Here, $N_\varphi$ represents the number of pairs falling into azimuthal bins centered at angle $\varphi$. For our broadband spectrum $n(\omega)$, the effective asymmetry $\langle R_B \rangle_{\mathrm{eff}}$ is averaged over the photon energy:
\begin{equation}
\langle R_B\rangle_{\mathrm{eff}} \approx \frac{\sin(2\beta)}{2\beta} \frac{\int d\omega\; n(\omega)\,\frac{\sigma_1(\omega)}{\sigma_0(\omega)}\,S_1(\omega)}{\int d\omega\; n(\omega)\,S_0(\omega)},
\end{equation}
where $2\beta$ is the azimuthal bin width. Based on our simulation, the generated average polarization $S_1 \approx 0.019$ translates to a measurable asymmetry $\langle R_B \rangle_{\mathrm{eff}} \approx 1.48 \times 10^{-3}$.

The statistical significance of the detection is determined by the total number of accumulated photons $N_\gamma$. To reach an $n_\sigma$ confidence level (e.g., $5\sigma$), the required photon number must satisfy:
\begin{equation}
N_\gamma = \frac{\pi n_\sigma^2}{4\eta_{\mathrm{eff}}\beta S_0\langle R_B\rangle_{\mathrm{eff}}^2}, \label{eq:Nreq_final}
\end{equation}
where $\eta_{\mathrm{eff}}$ is the effective pair-production conversion efficiency. 
Substituting our parameters into Eq.~(\ref{eq:Nreq_final}), the total photons required for a $5\sigma$ detection is $N_\gamma^{\mathrm{req}} \approx 2.59 \times 10^9$.  Further  details could be found in Supplymental Material\cite{SM}.

Since our baseline parameters ($a_0=125, E_e=3$~GeV, and $N_e = 10^8$) yield $N_\gamma \approx 1.21 \times 10^9$ photons per shot, the required number of shots is:
\begin{equation}
N_{\mathrm{shots}} = N_\gamma^{\mathrm{req}} / N_\gamma^{\mathrm{shot}} \approx 2.
\end{equation}
This demonstrates that a $5\sigma$ detection is feasible in just two laser shots at near-future multi-petawatt facilities. This high-efficiency scheme overcomes the low repetition rate of high-power lasers, providing a robust pathway for the first definitive observation of VB.
\begin{figure}[H]
  \centering
  \begin{overpic}[width=0.225\textwidth]{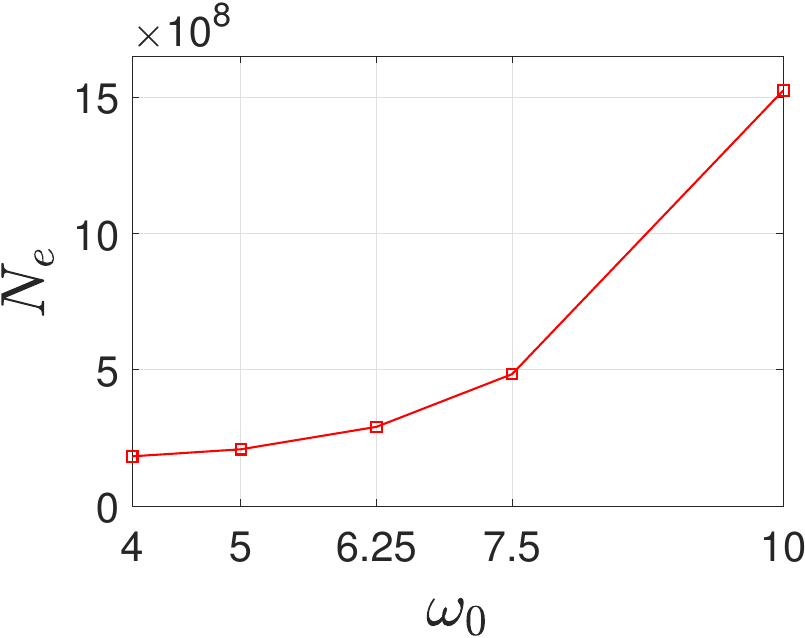}
      \put(0,75){\small (a)}
  \end{overpic}
  \hspace{2mm}
  \begin{overpic}[width=0.225\textwidth]{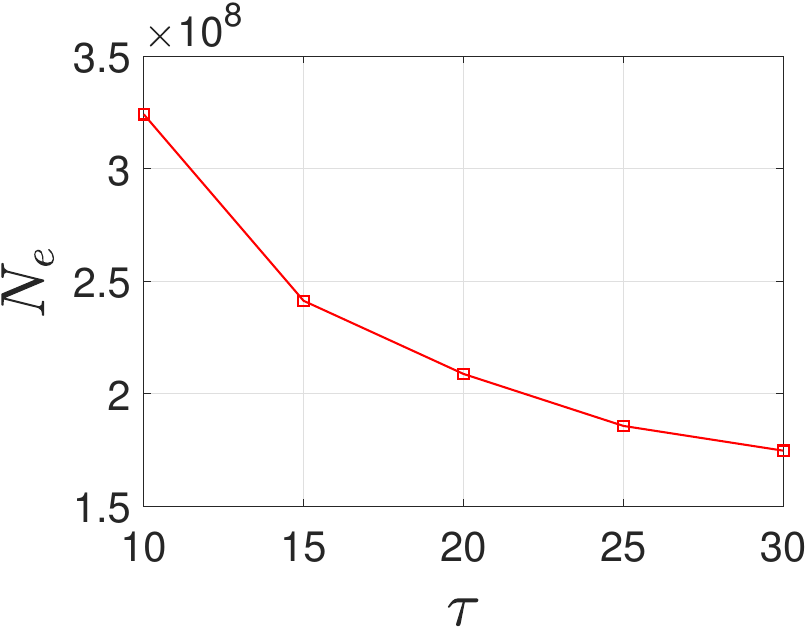}
      \put(0,75){\small (b)}
  \end{overpic}
  \vspace{2mm}
  \begin{overpic}[width=0.23\textwidth]{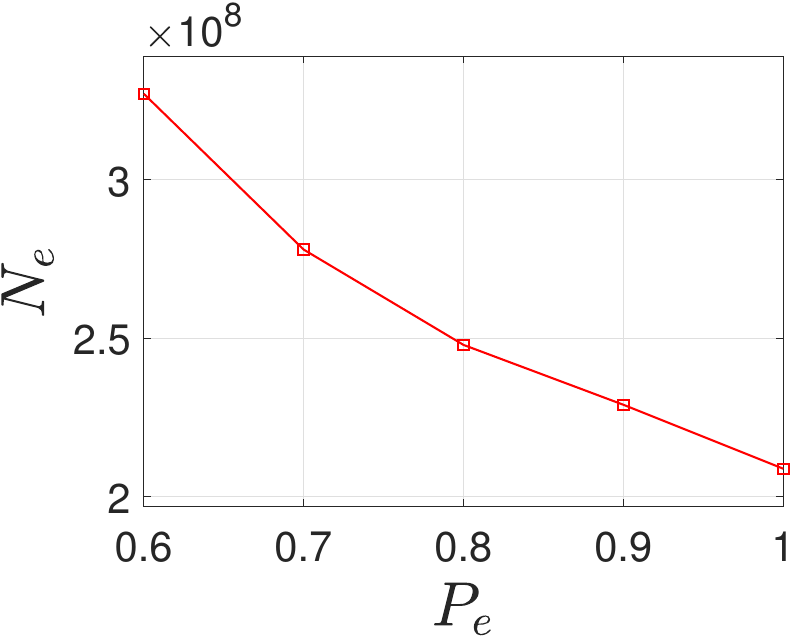}
      \put(0,75){\small (c)}
  \end{overpic}
  \hspace{2mm}
  \begin{overpic}[width=0.23\textwidth]{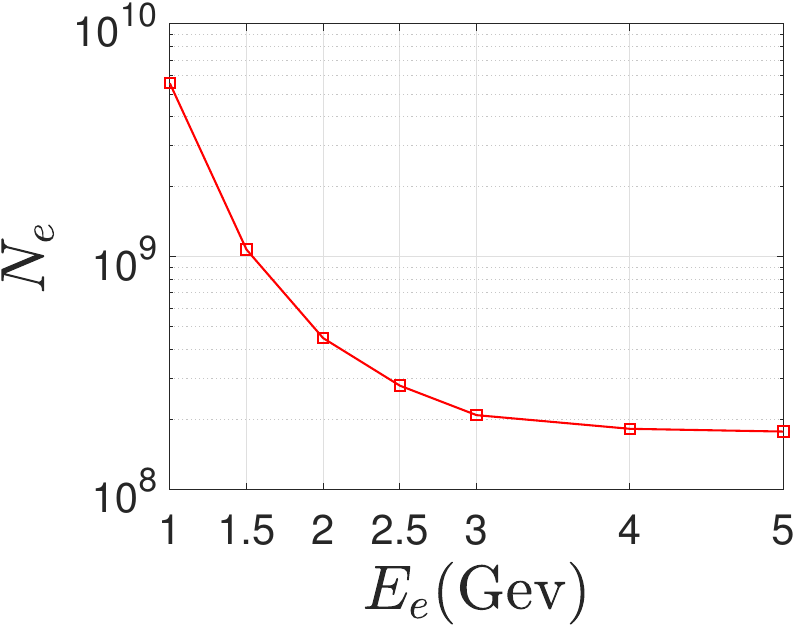}
      \put(-3,75){\small (d)}
  \end{overpic}
  \caption{Required initial electron number $N_e$ for a $5\sigma$ VB signal, plotted against (a) spot size $w_0$, (b) pulse duration $\tau_p$, (c) initial electron polarization, and (d) electron beam energy $E_e$.}
  \label{fig:example2}
\end{figure}

\begin{figure}[htbp]
  \centering
  \begin{overpic}[width=0.23\textwidth]{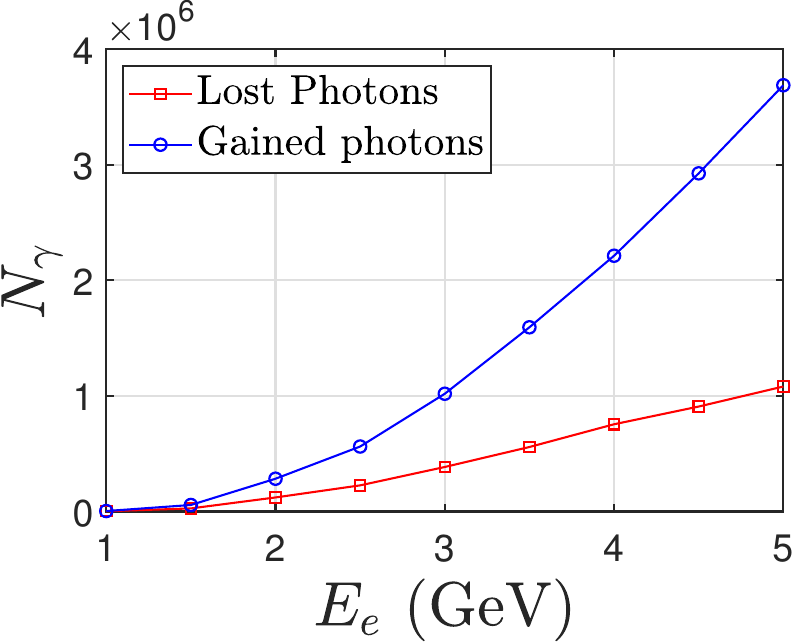}
      \put(0, 75){\small (a)}
      \label{fig:lost_num}
  \end{overpic}
  \hspace{2mm}
  \begin{overpic}[width=0.23\textwidth]{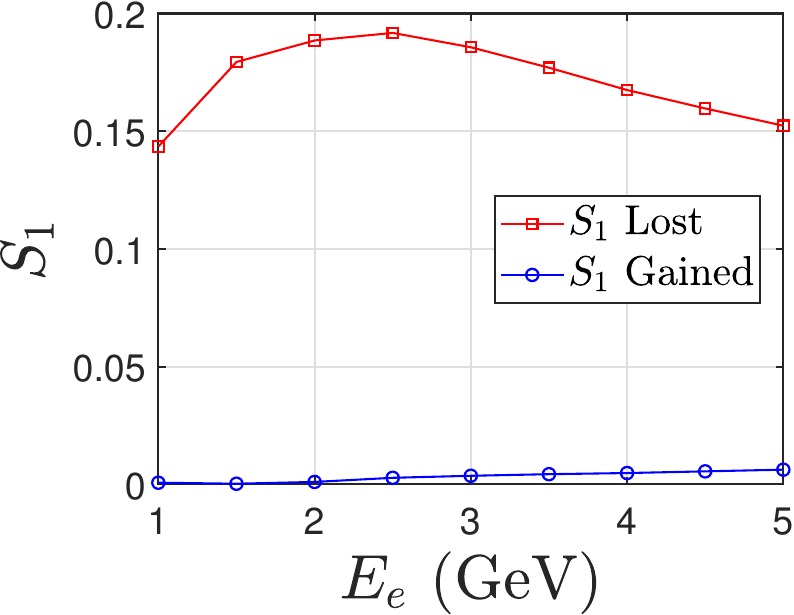}
      \put(0, 75){\small (b)}
      \label{fig:Lost_S1}
  \end{overpic}
\caption{Comparison of lost photons (absorbed) and gained photons (generated by electron pairs) vs. $E_e$. The exponential surge in unpolarized cascade photons at high energy dilutes the ensemble-averaged $S_1$ signal, leading to the $N_e$ plateau observed in Fig.~\ref{fig:example2}(d).}
 \label{fig:lost}
\end{figure}

Figure~\ref{fig:example2} shows the required $N_e$ as a function of experimental parameters under the constraint of a fixed 560~J laser energy. 
In Figure~\ref{fig:example2}(a), $N_e$ increases with $w_0$ because enlarging the spot size dilutes the peak intensity $a_0$. Since the VB rotation rate scales as $dS_1/dt \propto \chi_\gamma^{2/3}$, a lower $a_0$ leads to a drastically smaller $S_1$ signal, requiring more electrons.
Conversely, Figure~\ref{fig:example2}(b) shows that $N_e$ decreases for longer pulse durations $\tau_p$. With fixed laser energy ($a_0^2 \tau_p \approx \text{const}$), the accumulated signal scales as $S_1 \propto a_0^{2/3} \tau_p \propto a_0^{-4/3}$. Thus, reasonably lowering $a_0$ by stretching the pulse duration actually enhances the total signal by providing a much longer interaction time for the VB phase to accumulate, effectively reducing the required $N_e$.
While the initial electron polarization has a positive effect on the required $N_e$ [Figure~\ref{fig:example2}(c)],

Regarding the electron beam, Fig.~\ref{fig:example2}(c) confirms that increasing the initial electron polarization $P_e$ effectively reduces the required $N_e$. For electron beam energy, Figure~\ref{fig:example2}(d) shows a regime of diminishing returns.
As $E_e$ exceeds 3 GeV, the required $N_e$ plateaus. This is a direct consequence of the pair-production cascades. 
As shown in Figure~\ref{fig:lost}, while primary photons carrying the VB signal are absorbed at high $E_e$, the electron pairs generates a surge of secondary cascade photons. These unpolarized secondary photons dilute the ensemble-averaged $S_1$, counteracting the benefits of higher beam energy.

Therefore, the optimal $E_e$ must strike a balance: it must be high enough to ensure sufficient photon yield and $\chi_\gamma$, but low enough to prevent signal dilution via pair-production cascades. 
Our results confirm that $a_0 \sim 125$ and $E_e \sim 3$~GeV offer this optimal window, maximizing the birefringent signature while isolating it from cascade contamination.

\section{Conclusion}
In conclusion, we have proposed and numerically demonstrated a compact scheme for observing VB. 
By integrating $\gamma$-photon generation and probing into a single laser-electron interaction, our approach avoids key limitations of conventional pump-probe setups. This unified architecture eliminates the demanding requirements for femtosecond-level synchronization and external beam transport optics inherent to multi-beam experiments.
This guarantees a perfect spatiotemporal overlap and maximizes the accumulated phase shift, yielding a clear and robust signature: a generated linear polarization component ($S_{1} \approx 0.019$ for angular selected photons) that is the direct evidence of the induced ellipticity  within the birefringent quantum vacuum. 
We have identified an optimal operational window around $a_{0} \sim 125$ and $E_{e} \sim 3$ GeV, where the coherent birefringent signal is maximized while reducing the competing influence of VD. 
This trade-off highlights the importance of managing signal quality over mere photon quantity. 
The high photon yield of nonlinear Compton-scattering, combined with an analysis that integrates polarization information over the entire emitted spectrum, makes a 5$\sigma$ confidence-level detection feasible with as few as two laser shots at current multi-petawatt laser and accelerator facilities. 
This work therefore presents a robust and practical pathway toward a first definitive confirmation of VB. 
By enabling the simultaneous observation of both the dispersive (birefringence) and absorptive (dichroism) properties of the vacuum, this scheme opens up new path for experimentally studying the optical nature of the quantum vacuum in the strong-field limit.




\bibliographystyle{elsarticle-num}
\bibliography{bib/correction}

@article{Bragin2017,
  title = {High-Energy Vacuum Birefringence and Dichroism in an Ultrastrong Laser Field},
  author = {Bragin, Sergey and Meuren, Sebastian and Keitel, Christoph H. and Di Piazza, Antonino},
  year = {2017},
  month = dec,
  journal = {Physical Review Letters},
  volume = {119},
  number = {25},
  pages = {250403},
  issn = {0031-9007, 1079-7114},
  doi = {10.1103/PhysRevLett.119.250403},
  urldate = {2024-07-16},
  copyright = {https://link.aps.org/licenses/aps-default-license},
  langid = {english}
}

@article{Hubbell1980,
  title = {Pair, Triplet, and Total Atomic Cross Sections (and Mass Attenuation Coefficients) for 1 MeV-100 GeV Photons in Elements {\emph{Z}} =1 to 100},
  author = {Hubbell, J. H. and Gimm, H. A. and O/verbo/, I.},
  year = {1980},
  month = oct,
  journal = {Journal of Physical and Chemical Reference Data},
  volume = {9},
  number = {4},
  pages = {1023--1148},
  issn = {0047-2689, 1529-7845},
  doi = {10.1063/1.555629},
  urldate = {2025-03-27},
  langid = {english}
}

@article{aleksandrov2024vacuum,
  title={VACUUM BIREFRINGENCE AND DICHROISM IN A STRONG PLANE-WAVE BACKGROUND},
  author={Aleksandrov, IA and Shabaev, VM},
  journal={Zhurnal Eksperimentalnoi i Teoreticheskoi Fiziki (Journal of Experimental and Theoretical Physics)},
  volume={166},
  number={2},
  doi = {10.31857/S0044451024080042},
  pages={182--193},
  year={2024},
  publisher={The Russian Academy of Sciences}
}

@article{Dinu2014,
  title = {Vacuum Refractive Indices and Helicity Flip in Strong-Field QED},
  author = {Dinu, Victor and Heinzl, Tom and Ilderton, Anton and Marklund, Mattias and Torgrimsson, Greger},
  year = {2014},
  month = jun,
  journal = {Physical Review D},
  volume = {89},
  number = {12},
  primaryclass = {hep-ph, physics:hep-th},
  pages = {125003},
  issn = {1550-7998, 1550-2368},
  doi = {10.1103/PhysRevD.89.125003},
  urldate = {2024-07-30},
  archiveprefix = {arXiv},
  langid = {english}
}

@article{Groom:2000sm,
  title = {Passage of Particles through Matter: In Review of Particle Physics (RPP 2000)},
  author = {Groom, Donald E. and Klein, S.R.},
  year = 2000,
  journal = {The European Physical Journal C},
  volume = {15},
  number = {1-4},
  pages = {163--173},
  issn = {1434-6044, 1434-6052},
  doi = {10.1007/BF02683419},
  urldate = {2026-01-27},
  copyright = {http://www.springer.com/tdm},
  langid = {english}
}

@article{Alcorn:2004sb,
  title = {Basic Instrumentation for Hall A at Jefferson Lab},
  author = {Alcorn, J. and Anderson, B.D. and Aniol, K.A. and others},
  year = 2004,
  journal = {Nuclear Instruments and Methods in Physics Research Section A: Accelerators, Spectrometers, Detectors and Associated Equipment},
  volume = {522},
  number = {3},
  pages = {294--346},
  issn = {01689002},
  doi = {10.1016/j.nima.2003.11.415},
  urldate = {2026-01-27},
  langid = {english}
}

@article{Bernard2013a,
  title = {Polarimetry of Cosmic Gamma-Ray Sources above E+e- Pair Creation Threshold},
  author = {Bernard, Denis},
  year = 2013,
  month = nov,
  journal = {Nuclear Instruments and Methods in Physics Research Section A: Accelerators, Spectrometers, Detectors and Associated Equipment},
  volume = {729},
  eprint = {1307.3892},
  primaryclass = {astro-ph},
  pages = {765--780},
  issn = {01689002},
  doi = {10.1016/j.nima.2013.07.047},
  urldate = {2025-03-05},
  archiveprefix = {arXiv},
  langid = {english}
}

@article{aleksandrovLocallyconstantFieldApproximation2019,
  title = {Locally-Constant Field Approximation in Studies of Electron-Positron Pair Production in Strong External Fields},
  author = {Aleksandrov, I. A. and Plunien, G. and Shabaev, V. M.},
  year = {2019},
  month = jan,
  journal = {Physical Review D},
  volume = {99},
  number = {1},
  pages = {016020},
  issn = {2470-0010, 2470-0029},
  doi = {10.1103/PhysRevD.99.016020},
  urldate = {2024-08-29},
  langid = {english}
}

@article{bialynicka-birulaNonlinearEffectsQuantum1970,
  title = {Nonlinear Effects in Quantum Electrodynamics. Photon Propagation and Photon Splitting in an External Field},
  author = {{Bialynicka-Birula}, Z. and {Bialynicki-Birula}, Iwo},
  year = {1970},
  month = nov,
  journal = {Physical Review D},
  volume = {2},
  doi = {10.1103/PhysRevD.2.2341}
}

@article{Cadene2013,
  title = {Vacuum Magnetic Linear Birefringence Using Pulsed Fields: Status of the BMV Experiment},
  shorttitle = {Vacuum Magnetic Linear Birefringence Using Pulsed Fields},
  author = {Cad{\`e}ne, Agathe and Berceau, Paul and Fouch{\'e}, Mathilde and Battesti, R{\'e}my and Rizzo, Carlo},
  year = {2013},
  month = feb,
  journal = {The European Physical Journal D},
  volume = {68},
  doi = {10.1140/epjd/e2013-40725-9}
}

@article{dai2024fermionic,
  title={Fermionic signal of vacuum polarization in strong laser fields},
  author={Dai, Ya-Nan and Hatsagortsyan, Karen Z and Keitel, Christoph H and Chen, Yue-Yue},
  journal={Physical Review D},
  volume={110},
  number={1},
  pages={012008},
  year={2024},
  publisher={APS},
  doi = {10.1103/PhysRevD.110.012008}
}

@article{DellaValle2016,
  title = {The PVLAS Experiment: Measuring Vacuum Magnetic Birefringence and Dichroism with a Birefringent Fabry--Perot Cavity},
  shorttitle = {The PVLAS Experiment},
  author = {Della Valle, Federico and Ejlli, Aldo and Gastaldi, Ugo and Messineo, Giuseppe and Milotti, Edoardo and Pengo, Ruggero and Ruoso, Giuseppe and Zavattini, Guido},
  year = {2016},
  month = jan,
  journal = {The European Physical Journal C},
  volume = {76},
  number = {1},
  pages = {24},
  issn = {1434-6044, 1434-6052},
  doi = {10.1140/epjc/s10052-015-3869-8},
  urldate = {2024-07-26},
  langid = {english}
}

@incollection{dunne2005heisenberg,
  title={Heisenberg--Euler effective Lagrangians: basics and extensions},
  author={Dunne, Gerald V},
  booktitle={From Fields to Strings: Circumnavigating Theoretical Physics: Ian Kogan Memorial Collection (In 3 Volumes)},
  pages={445--522},
  year={2005},
  publisher={World Scientific},
  doi = {10.1142/5621}
}

@article{Ejlli2020,
  title = {The PVLAS Experiment: A 25 Year Effort to Measure Vacuum Magnetic Birefringence},
  shorttitle = {The PVLAS Experiment},
  author = {Ejlli, A. and Della Valle, F. and Gastaldi, U. and Messineo, G. and Pengo, R. and Ruoso, G. and Zavattini, G.},
  year = {2020},
  month = aug,
  journal = {Physics Reports},
  volume = {871},
  pages = {1--74},
  issn = {03701573},
  doi = {10.1016/j.physrep.2020.06.001},
  urldate = {2024-07-30},
  langid = {english}
}

@article{FolgerungenAusDiracschen1936,
  title = {Folgerungen aus der Diracschen Theorie des Positrons},
  author = {Heisenberg, Werner and Euler, Hans},
  year = {1936},
  journal = {Zeitschrift für Physik (Journal of Physics)},
  volume = {98},
  pages = {714--732},
  doi = {10.1007/BF01343663},
  langid = {german}
}

@article{Heinzl2006,
  title = {On the Observation of Vacuum Birefringence},
  author = {Heinzl, Thomas and Liesfeld, Bernhard and Amthor, Karl-Ulrich and Schwoerer, Heinz and Sauerbrey, Roland and Wipf, Andreas},
  year = {2006},
  month = {11},
  journal = {Optics Communications},
  volume = {267},
  number = {2},
  pages = {318--321},
  issn = {0030-4018},
  doi = {10.1016/j.optcom.2006.06.053}
}

@article{Karbstein2021,
  title = {Vacuum Birefringence at X-Ray Free-Electron Lasers},
  author = {Karbstein, Felix and Sundqvist, Chantal and Schulze, Kai S and Uschmann, Ingo and Gies, Holger and Paulus, Gerhard G},
  year = {2021},
  month = sep,
  journal = {New Journal of Physics},
  volume = {23},
  number = {9},
  pages = {095001},
  issn = {1367-2630},
  doi = {10.1088/1367-2630/ac1df4},
  urldate = {2024-10-14},
  langid = {english}
}

@article{torgrimsson2021loops,
  title={Loops and polarization in strong-field QED},
  author={Torgrimsson, Greger},
  journal={New Journal of Physics},
  volume={23},
  number={6},
  pages={065001},
  year={2021},
  publisher={IOP Publishing},
  doi = {10.1088/1367-2630/abf274}
}

@article{formanek2024signatures,
  title={Signatures of vacuum birefringence in low-power flying focus pulses},
  author={Formanek, Martin and Palastro, John P and Ramsey, Dillon and Weber, Stefan and Di Piazza, Antonino},
  journal={Physical Review D},
  volume={109},
  number={5},
  pages={056009},
  year={2024},
  publisher={APS},
  doi = {10.1103/PhysRevD.109.056009}
}

@article{Mei2010,
  title = {Axion Search with Q \& A Experiment},
  author = {Mei, Hsien-Hao and Ni, Wei-Tou and Chen, Sheng-Jui and Pan, Sheau-Shi},
  year = {2010},
  month = apr,
  journal = {Modern Physics Letters A},
  volume = {25},
  number = {11n12},
  primaryclass = {astro-ph, physics:hep-ex, physics:physics},
  pages = {983--993},
  issn = {0217-7323, 1793-6632},
  doi = {10.1142/S0217732310000149},
  urldate = {2024-08-01},
  archiveprefix = {arXiv},
  langid = {english}
}

@article{meurenPolarizationOperatorApproach2015,
  title = {Polarization Operator Approach to Pair Creation in Short Laser Pulses},
  author = {Meuren, Sebastian and Hatsagortsyan, Karen Z. and Keitel, Christoph H. and Di Piazza, Antonino},
  year = {2015},
  month = jan,
  journal = {Physical Review D},
  volume = {91},
  number = {1},
  primaryclass = {hep-ph},
  pages = {013009},
  issn = {1550-7998, 1550-2368},
  doi = {10.1103/PhysRevD.91.013009},
  urldate = {2024-08-07},
  langid = {english}
}

@article{nakamiya2017probing,
  title={Probing vacuum birefringence under a high-intensity laser field with gamma-ray polarimetry at the GeV scale},
  author={Nakamiya, Yoshihide and Homma, Kensuke},
  journal={Physical Review D},
  volume={96},
  number={5},
  pages={053002},
  year={2017},
  publisher={APS},
  doi = {10.1103/PhysRevD.96.053002}
}

@article{Schlenvoigt2016,
  title = {Detecting Vacuum Birefringence with X-Ray Free Electron Lasers and High-Power Optical Lasers: A Feasibility Study},
  shorttitle = {Detecting Vacuum Birefringence with X-Ray Free Electron Lasers and High-Power Optical Lasers},
  author = {Schlenvoigt, Hans-Peter and Heinzl, Tom and Schramm, Ulrich and Cowan, Thomas E and Sauerbrey, Roland},
  year = {2016},
  month = feb,
  journal = {Physica Scripta},
  volume = {91},
  number = {2},
  pages = {023010},
  issn = {0031-8949, 1402-4896},
  doi = {10.1088/0031-8949/91/2/023010},
  urldate = {2024-10-14},
  langid = {english}
}

@article{schwingerGaugeInvarianceVacuum1951,
  title = {On Gauge Invariance and Vacuum Polarization},
  author = {Schwinger, Julian},
  year = {1951},
  month = jun,
  journal = {Physical Review},
  volume = {82},
  number = {5},
  pages = {664--679},
  issn = {0031-899X},
  doi = {10.1103/PhysRev.82.664},
  urldate = {2024-08-22},
  copyright = {http://link.aps.org/licenses/aps-default-license},
  langid = {english}
}

@article{adler1971photon,
  title={Photon splitting and photon dispersion in a strong magnetic field},
  author={Adler, Stephen L},
  journal={Annals of Physics},
  volume={67},
  number={2},
  pages={599--647},
  year={1971},
  publisher={Elsevier},
  doi = {10.1016/0003-4916(71)90154-0}
}

@article{zavattini2007pvlas,
  title={PVLAS: probing vacuum with polarized light},
  author={Zavattini, Emilio and Zavattini, Guido and Ruoso, Giuseppe and Polacco, Erseo and Milotti, Edoardo and Karuza, Marin and Gastaldi, Ugo and Di Domenico, Giovanni and Della Valle, FEDERICO and Cimino, Roberto and others},
  journal={Nuclear Physics B: Proceedings Supplements},
  volume={164},
  pages={264--269},
  doi = {10.1103/PhysRev.82.664},
  year={2007},
  publisher={Elsevier}
}

@article{Mignani2017,
  title = {Evidence for Vacuum Birefringence from the First Optical-Polarimetry Measurement of the Isolated Neutron Star RX J1856.5-3754},
  author = {Mignani, R. P. and Testa, V. and Gonz{\'a}lez Caniulef, D. and Taverna, R. and Turolla, R. and Zane, S. and Wu, K.},
  year = {2017},
  month = feb,
  journal = {Monthly Notices of the Royal Astronomical Society},
  volume = {465},
  number = {1},
  pages = {492--500},
  issn = {0035-8711, 1365-2966},
  doi = {10.1093/mnras/stw2798},
  urldate = {2024-08-01},
  langid = {english}
}

@article{King2016,
  title = {Vacuum Birefringence in High-Energy Laser-Electron Collisions},
  author = {King, B. and Elkina, N.},
  year = {2016},
  month = dec,
  journal = {Physical Review A},
  volume = {94},
  number = {6},
  primaryclass = {hep-ph},
  pages = {062102},
  issn = {2469-9926, 2469-9934},
  doi = {10.1103/PhysRevA.94.062102},
  urldate = {2025-02-28},
  archiveprefix = {arXiv},
  langid = {english}
}

@article{McMaster1961,
  title = {Matrix Representation of Polarization},
  author = {McMaster, William H.},
  year = {1961},
  month = jan,
  journal = {Reviews of Modern Physics},
  volume = {33},
  number = {1},
  pages = {8--28},
  issn = {0034-6861},
  doi = {10.1103/RevModPhys.33.8},
  urldate = {2025-04-15},
  copyright = {http://link.aps.org/licenses/aps-default-license},
  langid = {english}
}

@article{Zhuang2023,
  title = {Laser-Driven Lepton Polarization in the Quantum Radiation-Dominated Reflection Regime},
  author = {Zhuang, Kai-Hong and Chen, Yue-Yue and Li, Yan-Fei and Hatsagortsyan, Karen Z. and Keitel, Christoph H.},
  year = {2023},
  month = aug,
  journal = {Physical Review D},
  volume = {108},
  number = {3},
  pages = {033001},
  issn = {2470-0010, 2470-0029},
  doi = {10.1103/PhysRevD.108.033001},
  urldate = {2025-03-01},
  langid = {english}
}

@article{wu2025,
  title={Achieving high polarization of photons emitted by unpolarized electrons in ultrastrong laser fields},
  author={Wu, Xian-Zhang and Li, Yan-Fei and Chen, Yue-Yue and Li, Yu-Tong},
  journal={Communications Physics},
  volume={8},
  number={1},
  pages={156},
  year={2025},
  publisher={Nature Publishing Group UK London},
  doi= {10.1038/s42005-025-02077-2}
}

@article{King2016a,
  title = {Measuring Vacuum Polarization with High-Power Lasers},
  author = {King, B. and Heinzl, T.},
  year = {2016},
  journal = {High Power Laser Science and Engineering},
  volume = {4},
  publisher = {Cambridge University Press (CUP)},
  issn = {2095-4719, 2052-3289},
  doi = {10.1017/hpl.2016.1},
  urldate = {2025-05-23},
  copyright = {http://creativecommons.org/licenses/by/4.0/},
  langid = {english}
}

@article{ahmadiniaz2025towards,
  title={Towards a vacuum birefringence experiment at the Helmholtz International Beamline for Extreme Fields (Letter of Intent of the BIREF@ HIBEF Collaboration)},
  author={Ahmadiniaz, Naser and B{\"a}htz, C and Benediktovitch, A and B{\"o}mer, C and Bocklage, L and Cowan, TE and Edwards, J and Evans, S and Vi{\~n}as, S Franchino and Gies, H and others},
  journal={High Power Laser Science and Engineering},
  volume={13},
  pages={e7},
  year={2025},
  publisher={Cambridge University Press},
  doi = {10.1017/hpl.2024.70}
}

@article{DiPiazza2012,
  title = {Extremely High-Intensity Laser Interactions with Fundamental Quantum Systems},
  author = {Di Piazza, A. and M{\"u}ller, C. and Hatsagortsyan, K. Z. and Keitel, C. H.},
  year = {2012},
  month = aug,
  journal = {Reviews of Modern Physics},
  volume = {84},
  number = {3},
  pages = {1177--1228},
  publisher = {American Physical Society (APS)},
  issn = {0034-6861, 1539-0756},
  doi = {10.1103/revmodphys.84.1177},
  urldate = {2025-05-23},
  copyright = {http://link.aps.org/licenses/aps-default-license},
  langid = {english}
}

@article{Ilderton2016,
  title = {Prospects for Studying Vacuum Polarisation Using Dipole and Synchrotron Radiation},
  author = {Ilderton, Anton and Marklund, Mattias},
  year = {2016},
  month = apr,
  journal = {Journal of Plasma Physics},
  volume = {82},
  number = {2},
  primaryclass = {hep-ph},
  issn = {0022-3778, 1469-7807},
  doi = {10.1017/S0022377816000192},
  urldate = {2025-07-09},
  archiveprefix = {arXiv},
  langid = {english}
}

@article{Chen2025,
  title = {Angular Modulation of Nonlinear Breit-Wheeler Yield by Vacuum Dichroism},
  author = {Chen, Jia-Ding and Dai, Ya-Nan and Zhuang, Kai-Hong and Jiang, Jing-Jing and Shen, Baifei and Chen, Yue-Yue},
  year = {2025},
  month = feb,
  journal = {Physical Review D},
  volume = {111},
  number = {3},
  pages = {036025},
  issn = {2470-0010, 2470-0029},
  doi = {10.1103/PhysRevD.111.036025},
  urldate = {2025-09-08},
  langid = {english}
}

@article{Ataman2018,
  title = {Vacuum Birefringence Detection in All-Optical Scenarios},
  author = {Ataman, Stefan},
  year = {2018},
  month = jun,
  journal = {Physical Review A},
  volume = {97},
  number = {6},
  primaryclass = {physics, physics:quant-ph},
  pages = {063811},
  issn = {2469-9926, 2469-9934},
  doi = {10.1103/PhysRevA.97.063811},
  urldate = {2024-07-26},
  archiveprefix = {arXiv},
  langid = {english}
}

@misc{SM,
  author = {{See Supplemental Material}},
  title  = {},
  note   = {at [URL will be inserted by publisher] for the method of calculating the number of required electrons to achieve sufficient signal photons.},
}

@article{Yoon,
author = {Jin Woo Yoon and Yeong Gyu Kim and Il Woo Choi and Jae Hee Sung and Hwang Woon Lee and Seong Ku Lee and Chang Hee Nam},
journal = {Optica},
number = {5},
pages = {630--635},
publisher = {Optica Publishing Group},
title = {Realization of laser intensity over 1023  W/cm2},
volume = {8},
month = {May},
year = {2021},
url = {https://opg.optica.org/optica/abstract.cfm?URI=optica-8-5-630},
doi = {10.1364/OPTICA.420520},

}

@article{Kim2008,
  title = {Drift-Free Femtosecond Timing Synchronization of Remote Optical and Microwave Sources},
  author = {Kim, Jungwon and Cox, Jonathan A. and Chen, Jian and K{\"a}rtner, Franz X.},
  year = 2008,
  month = dec,
  journal = {Nature Photonics},
  volume = {2},
  number = {12},
  pages = {733--736},
  issn = {1749-4885, 1749-4893},
  doi = {10.1038/nphoton.2008.225},
  urldate = {2025-10-23},
  copyright = {http://www.springer.com/tdm},
  langid = {english}
}

@article{Schulz2015,
  title = {Femtosecond All-Optical Synchronization of an X-Ray Free-Electron Laser},
  author = {Schulz, S. and Grgura{\v s}, I. and Behrens, C. and Bromberger, H. and Costello, J. T. and Czwalinna, M. K. and Felber, M. and Hoffmann, M. C. and Ilchen, M. and Liu, H. Y. and Mazza, T. and Meyer, M. and Pfeiffer, S. and Pr{\k e}dki, P. and Schefer, S. and Schmidt, C. and Wegner, U. and Schlarb, H. and Cavalieri, A. L.},
  year = 2015,
  month = jan,
  journal = {Nature Communications},
  volume = {6},
  number = {1},
  pages = {5938},
  issn = {2041-1723},
  doi = {10.1038/ncomms6938},
  urldate = {2025-10-23},
  langid = {english}
}

@article{Xin2018,
  title = {Ultra-Precise Timing and Synchronization for Large-Scale Scientific Instruments},
  author = {Xin, Ming and {\c S}afak, Kemal and K{\"a}rtner, Franz X.},
  year = 2018,
  month = dec,
  journal = {Optica},
  volume = {5},
  number = {12},
  pages = {1564},
  issn = {2334-2536},
  doi = {10.1364/OPTICA.5.001564},
  urldate = {2025-10-23},
  langid = {english}
}

@article{Zavattini2016,
  title = {A Polarisation Modulation Scheme for Measuring Vacuum Magnetic Birefringence with Static Fields},
  author = {Zavattini, G. and Della~Valle, F. and Ejlli, A. and Ruoso, G.},
  year = 2016,
  month = may,
  journal = {The European Physical Journal C},
  volume = {76},
  number = {5},
  pages = {294},
  issn = {1434-6044, 1434-6052},
  doi = {10.1140/epjc/s10052-016-4139-0},
  urldate = {2025-10-23},
  langid = {english}
}

@article{He2021,
  title = {Polarisation Optics for Biomedical and Clinical Applications: A Review},
  shorttitle = {Polarisation Optics for Biomedical and Clinical Applications},
  author = {He, Chao and He, Honghui and Chang, Jintao and Chen, Binguo and Ma, Hui and Booth, Martin J.},
  year = 2021,
  month = sep,
  journal = {Light: Science \& Applications},
  volume = {10},
  number = {1},
  pages = {194},
  issn = {2047-7538},
  doi = {10.1038/s41377-021-00639-x},
  urldate = {2025-10-23},
  langid = {english}
}

@article{Li2020,
  title = {Polarized Ultrashort Brilliant Multi-GeV \${\textbackslash}gamma\$-Rays via Single-Shot Laser-Electron Interaction},
  author = {Li, Yan-Fei and Shaisultanov, Rashid and Chen, Yue-Yue and Wan, Feng and Hatsagortsyan, Karen Z. and Keitel, Christoph H. and Li, Jian-Xing},
  year = 2020,
  month = jan,
  journal = {Physical Review Letters},
  volume = {124},
  number = {1},
  eprint = {1907.08877},
  primaryclass = {physics},
  pages = {014801},
  issn = {0031-9007, 1079-7114},
  doi = {10.1103/PhysRevLett.124.014801},
  urldate = {2024-07-17},
  archiveprefix = {arXiv},
  langid = {english}
}

@article{euler1935streuung,
  title={{\"U}ber die streuung von licht an licht nach der diracschen theorie},
  author={Euler, Hans and Kockel, Bernhard},
  journal={Naturwissenschaften},
  volume={23},
  number={15},
  pages={246--247},
  year={1935},
  publisher={Springer}
}

@article{mignani2016evidence,
  title={Evidence for vacuum birefringence from the first optical polarimetry measurement of the isolated neutron star RX J1856. 5- 3754},
  author={Mignani, RP and Testa, Vincenzo and Caniulef, D Gonz{\'a}lez and Taverna, Roberto and Turolla, Roberto and Zane, S and Wu, K},
  journal={Monthly Notices of the Royal Astronomical Society},
  pages={stw2798},
  year={2016},
  publisher={Oxford University Press}
}

@article{brandenburg2023report,
  title={Report on progress in physics: observation of the Breit--Wheeler process and vacuum birefringence in heavy-ion collisions},
  author={Brandenburg, James Daniel and Seger, Janet and Xu, Zhangbu and Zha, Wangmei},
  journal={Reports on Progress in Physics},
  volume={86},
  number={8},
  pages={083901},
  year={2023},
  publisher={IOP Publishing}
}

@article{Taverna:2022jgl,
  title = {Polarized X-Rays from a Magnetar},
  author = {Taverna, Roberto and Turolla, Roberto and Muleri, Fabio and others},
  year = 2022,
  month = may,
  journal = {Science},
  volume = {378},
  number = {6620},
  eprint = {2205.08898},
  primaryclass = {astro-ph.HE},
  pages = {646--650},
  issn = {0036-8075, 1095-9203},
  doi = {10.1126/science.add0080},
  urldate = {2025-12-26},
  archiveprefix = {arXiv},
  langid = {english}
}

@article{Salamin2002,
  title = {Electron Acceleration by a Tightly Focused Laser Beam},
  author = {Salamin, Yousef I. and Keitel, Christoph H.},
  year = 2002,
  month = feb,
  journal = {Physical Review Letters},
  volume = {88},
  number = {9},
  pages = {095005},
  issn = {0031-9007, 1079-7114},
  doi = {10.1103/PhysRevLett.88.095005},
  urldate = {2026-01-19},
  copyright = {http://link.aps.org/licenses/aps-default-license},
  langid = {english}
}

@book{landau4,
  author = {Berestetskii, V. B. and Lifshitz, E. M. and Pitaevskii, L. P.},
  title = {Quantum Electrodynamics},
  volume = {4},
  series = {Course of Theoretical Physics},
  edition = {2nd},
  publisher = {Butterworth-Heinemann},
  address = {Oxford},
  year = {1982}
}

@article{Tanaka2020,
  title = {Current Status and Highlights of the ELI-NP Research Program},
  author = {Tanaka, K. A. and Spohr, K. M. and Balabanski, D. L. and Balascuta, S. and Capponi, L. and Cernaianu, M. O. and Cuciuc, M. and Cucoanes, A. and Dancus, I. and Dhal, A. and Diaconescu, B. and Doria, D. and Ghenuche, P. and Ghita, D. G. and Kisyov, S. and Nastasa, V. and Ong, J. F. and Rotaru, F. and Sangwan, D. and S{\"o}derstr{\"o}m, P.-A. and Stutman, D. and Suliman, G. and Tesileanu, O. and Tudor, L. and Tsoneva, N. and Ur, C. A. and Ursescu, D. and Zamfir, N. V.},
  year = 2020,
  month = mar,
  journal = {Matter and Radiation at Extremes},
  volume = {5},
  number = {2},
  pages = {024402},
  issn = {2468-2047, 2468-080X},
  doi = {10.1063/1.5093535},
  urldate = {2026-02-03},
  langid = {english}
}

@article{ritus1985quantum,
  title={Quantum effects of the interaction of elementary particles with an intense electromagnetic field},
  author={Ritus, VI},
  journal={J. Sov. Laser Res.;(United States)},
  volume={6},
  number={5},
  year={1985}
}

@article{podszusHighenergyBehaviorStrongfield2019,
  title = {High-Energy Behavior of Strong-Field QED in an Intense Plane Wave},
  author = {Podszus, T. and Di Piazza, A.},
  year = 2019,
  month = apr,
  journal = {Physical Review D},
  volume = {99},
  number = {7},
  pages = {076004},
  issn = {2470-0010, 2470-0029},
  doi = {10.1103/PhysRevD.99.076004},
  urldate = {2024-08-26},
  langid = {english}
}

@article{DiPiazza:2018bfu,
  title = {Improved Local-Constant-Field Approximation for Strong-Field QED Codes},
  author = {Di Piazza, A. and Tamburini, M. and Meuren, S. and Keitel, C.H.},
  year = 2019,
  month = feb,
  journal = {Physical Review A},
  volume = {99},
  number = {2},
  eprint = {1811.05834},
  primaryclass = {hep-ph},
  pages = {022125},
  issn = {2469-9926, 2469-9934},
  doi = {10.1103/PhysRevA.99.022125},
  urldate = {2026-02-03},
  archiveprefix = {arXiv},
  langid = {english}
}

@book{baier1998electromagnetic,
  title={Electromagnetic processes at high energies in oriented single crystals},
  author={Baier, Vladimir N and Katkov, Valeri{\u\i} Mikha{\u\i}lovich and Strakhovenko, Vladimir Moiseevich},
  year={1998},
  publisher={World Scientific}
}

@article{yoon2021realization,
  title={Realization of laser intensity over $10^{23} {W/cm}$$^2$},
  author={Yoon, Jin Woo and Kim, Yeong Gyu and Choi, Il Woo and Sung, Jae Hee and Lee, Hwang Woon and Lee, Seong Ku and Nam, Chang Hee},
  journal={Optica},
  volume={8},
  number={5},
  pages={630--635},
  year={2021},
  publisher={Optical Society of America}
}

@article{Li2019,
  title = {Electron Polarimetry with Nonlinear Compton Scattering},
  author = {Li, Yan-Fei and Guo, Ren-Tong and Shaisultanov, Rashid and Hatsagortsyan, Karen Z. and Li, Jian-Xing},
  year = 2019,
  month = jul,
  journal = {Physical Review Applied},
  volume = {12},
  number = {1},
  pages = {014047},
  issn = {2331-7019},
  doi = {10.1103/PhysRevApplied.12.014047},
  urldate = {2025-02-28},
  langid = {english}
}

@article{Sevostyanov2021,
  title = {Total Yield of Electron-Positron Pairs Produced from Vacuum in Strong Electromagnetic Fields: Validity of the Locally Constant Field Approximation},
  shorttitle = {Total Yield of Electron-Positron Pairs Produced from Vacuum in Strong Electromagnetic Fields},
  author = {Sevostyanov, D. G. and Aleksandrov, I. A. and Plunien, G. and Shabaev, V. M.},
  year = 2021,
  month = oct,
  journal = {Physical Review D},
  volume = {104},
  number = {7},
  pages = {076014},
  issn = {2470-0010, 2470-0029},
  doi = {10.1103/PhysRevD.104.076014},
  urldate = {2024-08-29},
  langid = {english}
}

@misc{Aleksandrov2023,
  title = {Vacuum Birefringence and Dichroism in a Strong Plane-Wave Background},
  author = {Aleksandrov, I. A. and Shabaev, V. M.},
  year = 2023,
  month = mar,
  number = {arXiv:2303.16273},
  eprint = {2303.16273},
  primaryclass = {hep-ph},
  publisher = {arXiv},
  urldate = {2024-08-01},
  archiveprefix = {arXiv},
  langid = {english}
}

@article{Mosman:2021vua,
  title = {Vacuum Birefringence and Diffraction at an X-Ray Free-Electron Laser: From Analytical Estimates to Optimal Parameters},
  shorttitle = {Vacuum Birefringence and Diffraction at an X-Ray Free-Electron Laser},
  author = {Mosman, Elena A. and Karbstein, Felix},
  year = 2021,
  month = jul,
  journal = {Physical Review D},
  volume = {104},
  number = {1},
  eprint = {2104.05103},
  primaryclass = {hep-ph},
  pages = {013006},
  issn = {2470-0010, 2470-0029},
  doi = {10.1103/PhysRevD.104.013006},
  urldate = {2026-02-19},
  archiveprefix = {arXiv},
  langid = {english}
}

@misc{Lv2024,
  title = {Generation of High-Brilliance Polarized \$\textbackslash gamma\$-Rays via Vacuum Dichroism-Assisted Vacuum Birefringence},
  author = {Lv, Chong and Wan, Feng and Salamin, Yousef I. and Zhao, Qian and Ababekri, Mamutjan and Xu, Ruirui and Li, Jian-Xing},
  year = 2024,
  month = apr,
  number = {arXiv:2401.14075},
  eprint = {2401.14075},
  primaryclass = {physics},
  publisher = {arXiv},
  urldate = {2024-07-28},
  archiveprefix = {arXiv},
  langid = {english}
}

@article{Wistisen2013,
  title = {Vacuum Birefringence by Compton Backscattering through a Strong Field},
  author = {Wistisen, Tobias N. and Uggerh{\o}j, Ulrik I.},
  year = 2013,
  month = sep,
  journal = {Physical Review D},
  volume = {88},
  number = {5},
  pages = {053009},
  issn = {1550-7998, 1550-2368},
  doi = {10.1103/PhysRevD.88.053009},
  urldate = {2025-08-15},
  copyright = {http://link.aps.org/licenses/aps-default-license},
  langid = {english}
}

@article{thomas1926motion,
  title={The motion of the spinning electron},
  author={Thomas, Llewellyn H},
  journal={Nature},
  volume={117},
  number={2945},
  pages={514--514},
  year={1926},
  publisher={Nature Publishing Group UK London}
}

@article{thomas1927kinematics,
  title={I. The kinematics of an electron with an axis},
  author={Thomas, Llewellyn Hilleth},
  journal={The London, Edinburgh, and Dublin Philosophical Magazine and Journal of Science},
  volume={3},
  number={13},
  pages={1--22},
  year={1927},
  publisher={Taylor \& Francis}
}

@article{bargmann1959precession,
  title={Precession of the polarization of particles moving in a homogeneous electromagnetic field},
  author={Bargmann, Valentine and Michel, Louis and Telegdi, VL},
  journal={Physical Review Letters},
  volume={2},
  number={10},
  pages={435},
  year={1959},
  publisher={APS}
}

@article{walser2002spin,
  title={Spin and radiation in intense laser fields},
  author={Walser, MW and Urbach, David J and Hatsagortsyan, Karen Zaven and Hu, SX and Keitel, Christoph H},
  journal={Physical Review A},
  volume={65},
  number={4},
  pages={043410},
  year={2002},
  publisher={APS}
}

@article{li2020production,
  title={Production of highly polarized positron beams via helicity transfer from polarized electrons in a strong laser field},
  author={Li, Yan-Fei and Chen, Yue-Yue and Wang, Wei-Min and Hu, Hua-Si},
  journal={Physical Review Letters},
  volume={125},
  number={4},
  pages={044802},
  year={2020},
  publisher={APS}
}

@article{ivanov2005complete,
  title={Complete description of polarization effects in e+ e-pair productionby a photon in the field of a strong laser wave},
  author={Ivanov, D Yu and Kotkin, GL and Serbo, VG},
  journal={The European Physical Journal C-Particles and Fields},
  volume={40},
  number={1},
  pages={27--40},
  year={2005},
  publisher={Springer}
}

@article{Gonsalves2019,
  title={Petawatt laser guiding and electron beam acceleration to 8 GeV in a laser-heated capillary discharge waveguide},
  author={Gonsalves, AJ and Nakamura, K and Daniels, J and Benedetti, C and Pieronek, C and De Raadt, TCH and Steinke, S and Bin, JH and Bulanov, SS and Van Tilborg, J and others},
  journal={Physical review letters},
  volume={122},
  number={8},
  pages={084801},
  year={2019},
  publisher={APS}
}

@article{tanaka2020current,
  title={Current status and highlights of the ELI-NP research program},
  author={Tanaka, KA and Spohr, KM and Balabanski, DL and Balascuta, S and Capponi, L and Cernaianu, MO and Cuciuc, M and Cucoanes, A and Dancus, I and Dhal, A and others},
  journal={Matter and Radiation at Extremes},
  volume={5},
  number={2},
  year={2020},
  publisher={AIP Publishing}
}

\end{document}


\begin{frontmatter}

\title{Supplemental Material for "Probing Vacuum Birefringence in an Ultrastrong Laser Field via High-energy Gamma-ray Polarimetry"}

\end{frontmatter}

\setcounter{equation}{0}
\setcounter{figure}{0}
\setcounter{table}{0}
\setcounter{page}{1}
\setcounter{section}{0}
\makeatletter
\renewcommand{\theequation}{S\arabic{equation}}
\renewcommand{\thefigure}{S\arabic{figure}}
\renewcommand{\thetable}{S\arabic{table}}
\makeatother

\section{Statistical Significance and Confidence Level Evaluation}

To thoroughly evaluate the feasibility of achieving a $5\sigma$ confidence-level detection in our setup, we present a detailed statistical derivation. We follow the formalism of Bragin \textit{et al.} \cite{Bragin2017} for the polarization asymmetry and extend it to rigorously account for the broadband energy distribution of the probe photons in our experiment.

\subsection{1. Statistical framework for a monochromatic beam}

In the polarimetry scheme, the observable is the asymmetry $R_B$ defined from the azimuthal distribution of electron-positron pairs produced in a converter:
\begin{equation}
R_B = \frac{(N_{\pi/4}+N_{5\pi/4})-(N_{3\pi/4}+N_{7\pi/4})}{(N_{\pi/4}+N_{5\pi/4})+(N_{3\pi/4}+N_{7\pi/4})}.
\end{equation}
For a quasi-monochromatic beam, the expectation value of $R_B$ is given by \cite{Bragin2017}:
\begin{equation}
\langle R_B \rangle = \frac{\sin(2\beta)}{2\beta}\,\frac{\sigma_1}{\sigma_0}\,\frac{S_1}{S_0}, \label{eq:RB_mono}
\end{equation}
where $\beta$ is the half-width of the azimuthal bins, $\sigma_0,\sigma_1$ are the pair-production cross-section coefficients, and $S_0,S_1$ are the Stokes parameters of the photon beam.

The statistical uncertainty of $R_B$ follows from error propagation applied to the multinomial distribution of the detected pairs. Assuming each photon produces a pair with probabilities $p_A$ and $p_B$ to fall into the two orthogonal azimuthal bins used for $R_B$, one obtains the standard deviation:
\begin{equation}
\langle (\Delta R_B)^2\rangle = \frac{1-\langle R_B\rangle^2}{N_\gamma(p_A+p_B)}.
\end{equation}
In the weak-signal regime ($\langle R_B\rangle^2 \ll 1$), to claim a detection with a significance of $n_\sigma$ standard deviations ($\langle R_B\rangle = n_\sigma\,\Delta R_B$), the required number of photons $N_\gamma$ is:
\begin{equation}
N_\gamma = \frac{n_\sigma^2}{\langle R_B\rangle^2\,(p_A+p_B)}. \label{eq:N_gamma_general}
\end{equation}
By integrating the differential cross section over the azimuthal intervals of width $2\beta$, one finds $p_A+p_B = \frac{4\beta}{\pi}\eta S_0$, where $\eta = n_z l\,\sigma_0$ is the conversion efficiency. Equation (\ref{eq:N_gamma_general}) then becomes:
\begin{equation}
N_\gamma = \frac{\pi n_\sigma^2}{4\eta\beta S_0\langle R_B\rangle^2}. \label{eq:Nreq_mono}
\end{equation}

\subsection{2. Extension to a broadband photon spectrum}

\begin{figure}[H]
  \centering
  \begin{subfigure}[b]{0.235\textwidth}
      \centering
      \includegraphics[width=\textwidth]{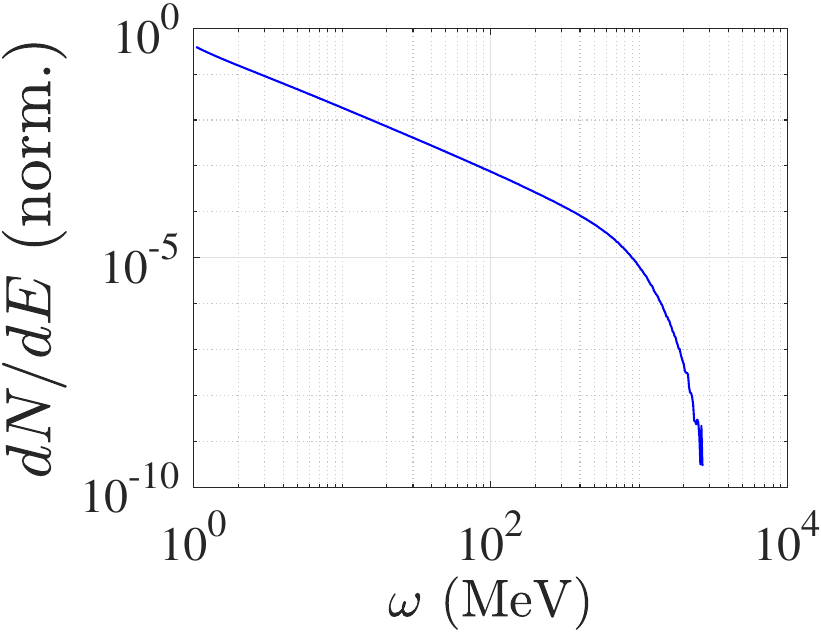}
      \put(-125, 90){\small (a)}
  \end{subfigure}
  \begin{subfigure}[b]{0.235\textwidth}
      \centering
      \includegraphics[width=\textwidth]{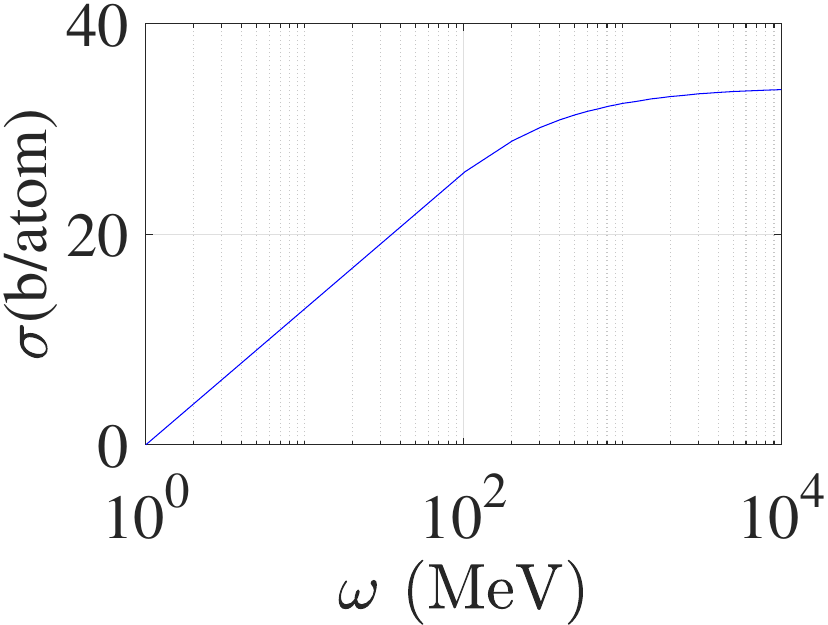}
      \put(-125, 90){\small (b)}
  \end{subfigure}
  \caption{(a) Simulated photon energy spectrum $n(\omega)$ and corresponding $S_1(\omega)$. (b) Photon pair production cross sections $\sigma_0(\omega)$ on a Tungsten (W) target.}
  \label{fig:integral}
\end{figure}

In our setup, the probe photons exhibit a broad energy spectrum $n(\omega)$ (see Fig.~\ref{fig:integral}(a)). Consequently, both the Stokes parameters and the cross-section ratio $\sigma_1/\sigma_0$ depend on photon energy. The effective asymmetry must be averaged over the spectrum:
\begin{equation}
\langle R_B\rangle_{\mathrm{eff}} = \frac{\sin 2\beta}{2\beta}\,
\frac{\int d\omega\; n(\omega)\,\frac{\sigma_1(\omega)}{\sigma_0(\omega)}\,S_1(\omega)}
{\int d\omega\; n(\omega)\,S_0(\omega)}. \label{eq:RB_eff}
\end{equation}

For our baseline parameters, the initial beam has $S_1 \approx 0$, and vacuum birefringence generates a small $S_1$ component. From our simulation (discarding photons below 1.02 MeV), we obtain the energy-resolved Stokes parameter $S_1(\omega)$ and the photon spectrum $n(\omega)$. The cross-sections $\sigma_0(\omega)$ for a tungsten converter ($Z=74$) are extracted from the Hubbell \cite{Hubbell1980} and are shown in Fig.~\ref{fig:integral}(b).

Performing the numerical integration yields:
\begin{equation}
\langle R_B\rangle_{\mathrm{eff}} \approx 1.481 \times 10^{-3}.
\end{equation}
This value is rigorously consistent with the energy-averaged Stokes parameter and the approximate monochromatic cross-section ratio once the spectral weights are taken into account. 

The effective conversion efficiency also requires spectral averaging:
\begin{equation}
\eta_{\mathrm{eff}} = n_z l\,\frac{\int d\omega\; n(\omega)\,\sigma_0(\omega)}{\int d\omega\; n(\omega)} = n_z l\,\langle\sigma_0\rangle.
\end{equation}
By integrating the pair-production cross-sections over our specific photon spectrum for a chosen 10 mRL Tungsten converter thickness, the effective conversion efficiency is calculated to be $\eta_{\mathrm{eff}} \approx 6.005 \times 10^{-3}$. Substituting $\langle R_B\rangle_{\mathrm{eff}} = 1.481\times10^{-3}$, $\eta_{\mathrm{eff}} = 6.005\times 10^{-3}$, $\beta = 33^\circ$ (i.e., $0.576$ rad), $S_0\approx 1$, and $n_\sigma = 5$ into Eq.~(\ref{eq:Nreq_mono}) gives:
\begin{equation}
N_\gamma^{\mathrm{req}} = \frac{\pi \times 5^2}{4\times 6.005 \times 10^{-3} \times 0.576 \times 1 \times (1.481 \times 10^{-3})^{2}} \approx 2.587\times10^{9}.
\end{equation}
With a per-shot photon yield of $N_\gamma^{\mathrm{shot}} \approx 1.21\times10^{9}$ (for $N_e = 10^8$ initial electrons), the number of required shots is:
\begin{equation}
N_{\mathrm{shots}} = \frac{N_\gamma^{\mathrm{req}}}{N_\gamma^{\mathrm{shot}}} \approx 2.
\end{equation}
This demonstrates that the broadband high-flux scheme can achieve a $5\sigma$ confidence level in just a few shots.

\section{Experimental Setup and Detection Strategy}

\subsection{Signal Optimization via Energy Selection (Optional)}

While the baseline analysis in the main text demonstrates that a $5\sigma$ significance can be readily achieved by integrating over the generated broadband spectrum, the statistical weight of the vacuum birefringence signal is not uniformly distributed across all energies. Based on Eq.~(\ref{eq:Nreq_mono}), the required statistics scale as $N_\gamma \propto S_1^{-2} \sigma^{-1}$. To identify the most efficient energy range, we define a statistical sensitivity metric $S \propto N_\gamma S_1^2 \sigma$, which directly determines the confidence level of the signal.

\begin{figure}[htbp]
  \centering
  \includegraphics[scale=0.4]{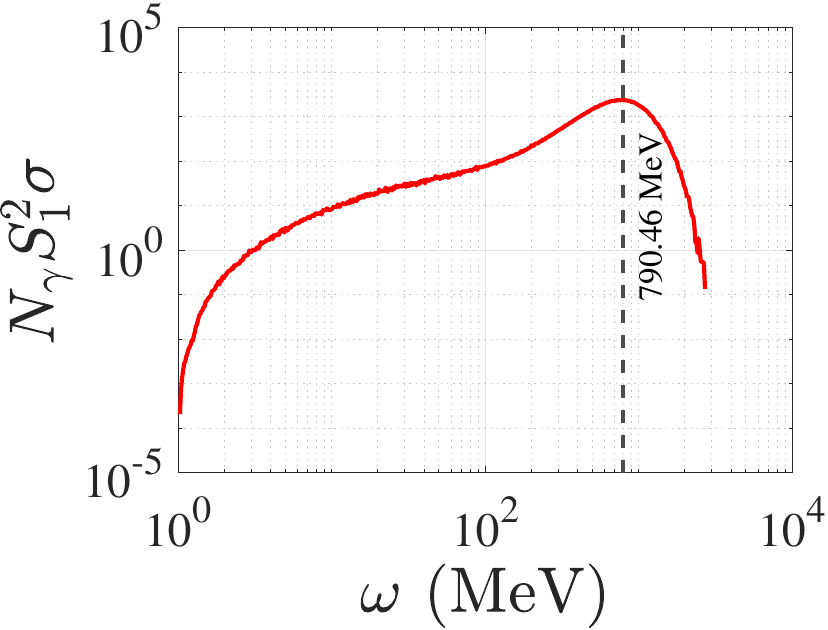}
  \caption{\label{fig:threshold} Statistical sensitivity metric $N_\gamma S_1^2 \sigma$ as a function of photon energy $\omega_\gamma$. The metric reaches its maximum near $\sim 790$ MeV.}
\end{figure}

As shown in Fig.~\ref{fig:threshold}, this sensitivity metric is negligible below 100~MeV but peaks strongly around $\omega_\gamma \approx 790$~MeV. Therefore, an optional but highly effective experimental strategy to mitigate the high flux of sub-GeV background photons—which might otherwise saturate sensitive detectors—is to employ a magnetic spectrometer immediately after the conversion target. 

It is crucial to note that the magnetic field acts on charged particles, not the photons themselves. By tuning the spectrometer to selectively deflect the $e^+e^-$ pairs generated by photons in the $750-850$ MeV window, one can spatially separate the high-value signal pairs from the dense, straight-line background of unconverted low-energy photons and broadly dispersed low-energy pairs. This spatial separation directs only the kinematically selected signal pairs onto the Image Plates, thereby significantly optimizing the experimental signal-to-noise ratio without distorting the intrinsic azimuthal polarization signature.
\subsection{Beam Parameters and Pointing Stability}

A vital aspect of experimental feasibility is the stability and physical size of the electron beam relative to the laser focal spot. We conservatively adopt an electron beam size of $\sigma_e = 2\,\mu\text{m}$. 
Because the electrons are distributed across the nonparaxial Gaussian profile of the laser ($w_0 = 5\,\mu\text{m}$), the ensemble-averaged quantum parameter $\chi$ is slightly lower than for an ideal point-like beam. 
However, this geometric reduction is easily compensated by a modest increase in the total electron bunch charge (to $N_e \approx 2.247 \times 10^8$, corresponding to a highly feasible $5\%$ influence).

\begin{figure}[htbp]
  \centering
  \includegraphics[scale=0.5]{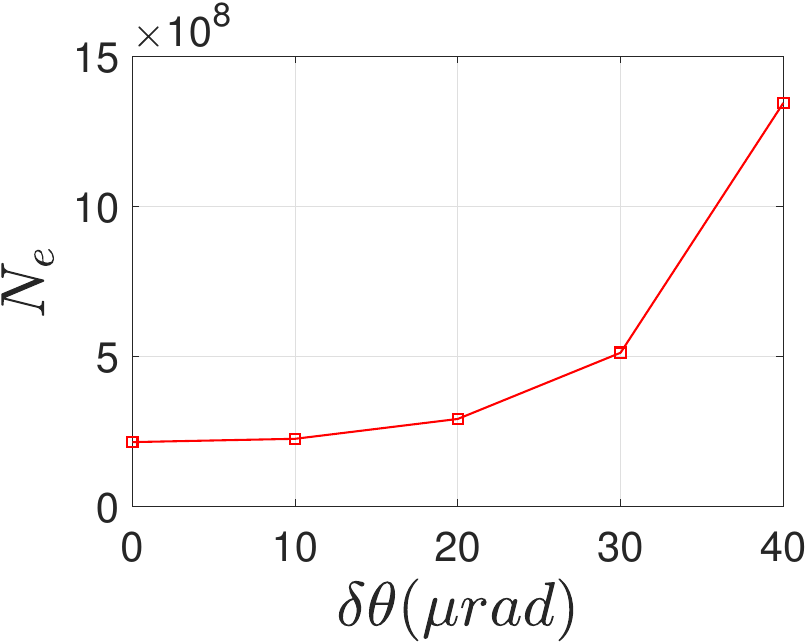}
  \caption{\label{fig:jitter} Number of initial electrons $N_e$ required for a $5\sigma$ significance detection under varying angular beam jitter conditions.}
\end{figure}

Furthermore, the setup is highly robust against typical spatial jitter. For an interaction distance of $10\,\text{cm}$, a realistic angular jitter of $\delta\theta = 30\,\mu\text{rad}$ results in a $3\,\mu\text{m}$ transverse displacement at the focus. Since the laser waist $w_0 = 5\,\mu\text{m}$ remains larger than both the beam size and the jitter displacement, the interaction reliably occurs within the high-intensity core of the laser pulse. As shown in Fig.~\ref{fig:jitter}, even with increasing jitter, the required number of electrons to maintain a $5\sigma$ significance remains well within the capabilities of current conventional accelerator facilities.

\subsection{Background and Systematic Uncertainties}

\textbf{Residual Gas Background:} Under typical high-vacuum conditions ($< 10^{-6}$ Torr) within the interaction chamber, the background gamma-ray yield originating from electron bremsstrahlung with residual gas molecules is estimated to be multiple orders of magnitude lower than the signal generated via nonlinear Compton scattering. Furthermore, bremsstrahlung photons are predominantly concentrated in the very low-energy regime, posing a negligible impact on the detection of the GeV-scale polarization signatures.

\textbf{Multiple Coulomb Scattering (MCS):} The linear polarization of the incident photons ($S_1$) is determined by observing the 'X-shaped' azimuthal distribution of the $e^+e^-$ pairs. A primary physical source of systematic uncertainty is the blurring of this spatial distribution due to MCS as pairs exit the converter foil. For $\sim 400$ MeV electrons/positrons traversing a 10 mRL Tungsten target ($\sim 35 \mu\text{m}$), the RMS scattering angle is calculated as \cite{Groom:2000sm}:
\begin{equation}
\theta_0 = \frac{13.6 \text{ MeV}}{\beta cp} z \sqrt{x/X_0} [1 + 0.038 \ln(x/X_0)] \approx 3 \text{ mrad}.
\end{equation}
Given that the intensity profile is typically recorded at a characteristic polar angle (beam spot size) of $\theta_{\text{obs}} \approx 10$ mrad, the effective azimuthal resolution is $\sigma_\phi \approx \theta_0 / \theta_{\text{obs}} \approx 0.3$. Based on the angular dilution model \cite{Bernard2013a}, this results in a dilution factor $D = \exp(-2\sigma_\phi^2) \approx 0.84$. This means approximately 84\% of the intrinsic polarization signal is preserved. Because the scattering-induced angular migration ($\sim 3$ mrad) is significantly smaller than the macroscopic azimuthal span of the X-shape quadrants, the structural integrity of the polarization signature remains robust against MCS blurring.

\textbf{Instrumental Asymmetries:} Finally, instrumental defects such as magnetic field inhomogeneities or detector misalignments could potentially induce a 'fake' azimuthal modulation. However, in modern experimental setups, detector relative alignments can be routinely maintained to a precision better than 1 mrad \cite{Alcorn:2004sb}. Any geometric distortion from the apparatus is therefore sub-dominant compared to the physical scattering scale ($\theta_0 \approx 3$ mrad), ensuring that the observed excess of particle density in the diagonal quadrants is a genuine QED effect of the photon polarization state rather than an artifact of experimental bias.

\nocite{*}
\bibliography{bib/SM_correction}
\bibliographystyle{elsarticle-num}